\newcommand{\etal}{\emph{et al.}}
\newcommand{\eg}{\emph{e.g.,}}
\newcommand{\ie}{\emph{i.e.}}

\newcommand{\clnote}[1]{{#1}}
\newcommand{\ProjectName}{{\small\textsc{Asteria}}}  

\newcommand{\toolname}[1]{{\small\textsc{#1}}}

\newcommand{\siamese}{\textit{Siamese Network}}
\newcommand{\diaphora}{{\textit{Diaphora}}}
\newcommand{\gemini}{{\textit{Gemini}}}
\newcommand{\ysgbf}[1]{{\small{\textbf{#1}}}}

\newcommand{\vs}{\emph{vs.}}
\newcommand{\del}[1]{}
\newcommand{\ProjectUrl}{\textit{https://github.com/Asteria-BCSD/Asteria}}
\newcommand{\nodename}[1]{\textit{#1}}

\documentclass[conference]{IEEEtran}
\AtBeginDocument{%
  \providecommand\BibTeX{{%
    \normalfont B\kern-0.5em{\scshape i\kern-0.25em b}\kern-0.8em\TeX}}}

\usepackage[labelformat=simple]{subcaption}

\usepackage{multirow}
\usepackage{subcaption}
\usepackage{graphicx}
\usepackage{xcolor}
\usepackage{amsmath}
\usepackage{array}
\usepackage{enumitem}
\usepackage{dashrule}
\usepackage{float}
\usepackage[utf8]{inputenc}
\usepackage{pifont}
\usepackage{hyperref}
\usepackage{cleveref}
\usepackage{colortbl}
\usepackage{tabulary}
\usepackage{tabularx}
\crefname{section}{§}{§§}

\begin{document}

\title{Asteria: Deep Learning-based AST-Encoding for Cross-platform Binary Code Similarity Detection}


\author{\IEEEauthorblockN{Shouguo Yang\IEEEauthorrefmark{1}\IEEEauthorrefmark{2}, Long Cheng\IEEEauthorrefmark{3},
Yicheng Zeng\IEEEauthorrefmark{1}\IEEEauthorrefmark{2}, Zhe Lang\IEEEauthorrefmark{1}\IEEEauthorrefmark{2}, Hongsong Zhu\IEEEauthorrefmark{1}\IEEEauthorrefmark{2}, Zhiqiang Shi\IEEEauthorrefmark{1}\IEEEauthorrefmark{2}},
\IEEEauthorblockA{\IEEEauthorrefmark{1}\textit{Institute of Information Engineering, Chinese Academy of Sciences, Beijing, China} \\
\IEEEauthorrefmark{2}\textit{School of Cyber Security, University of Chinese Academy of Sciences, Beijing China}\\
\IEEEauthorrefmark{3}\textit{School of Computing, Clemson University, USA}\\
\{yangshouguo, zengyicheng, langzhe, zhuhongsong, shizhiqiang\}@iie.ac.cn\\lcheng2@clemson.edu}
}
\maketitle

%

\begin{abstract}
Binary code similarity detection is a fundamental technique for many security applications such as vulnerability search, patch analysis, and malware detection. 
There is an increasing need to detect similar code for vulnerability search across architectures with the increase of critical vulnerabilities in IoT devices. 
The variety of IoT hardware architectures and software platforms requires to capture semantic equivalence of code fragments in the similarity detection. 
However, existing approaches are insufficient in capturing the semantic similarity.
We notice that the abstract syntax tree (AST) of a function contains rich semantic information.
Inspired by successful applications of natural language processing technologies in sentence semantic understanding, we propose a deep learning-based AST-encoding method, named {\small\textsc{Asteria}}, to measure the semantic equivalence of functions in different platforms.
Our method leverages the Tree-LSTM network to learn the semantic representation of a function from its AST.
Then the similarity detection can be conducted efficiently and accurately by measuring the similarity between two representation vectors.
We have implemented an open-source prototype of \ProjectName{}. 
The Tree-LSTM model is trained on a dataset with 1,022,616 function pairs and evaluated on a dataset with 95,078 function pairs.
Evaluation results show that our method outperforms the AST-based tool \diaphora{} and the-state-of-art method \gemini{} by large margins with respect to the binary similarity detection.
And our method is several orders of magnitude faster than \diaphora{} and \gemini{} for the similarity calculation.
In the application of vulnerability search, our tool successfully identified 75 vulnerable functions in 5,979 IoT firmware images. 
\end{abstract}

\section{Introduction} \label{sec:intro}
Over recent years, we have witnessed the rapid development and deployment of the Internet of Things (IoT). 
{However, the pressure of time to market of IoT development increasingly raises security and privacy concerns~\cite{Wurm:DAC:2016}. } 
Firmware of IoT devices could contain vulnerabilities, which have already caused destructive attacks~\cite{Hernandez:Firmware:2019}. 
IoT firmware security analysis is considered an effective approach to ensuring the security of IoT devices~\cite{Costin:2014:Security}.

Due to the lack of source code, analyzing binary code has naturally become an important means of firmware security analysis.
On the other hand, code is often reused to facilitate fast software development. 
Unfortunately, the reused code such as a third-party library could cause the same vulnerability to spread among different vendors and different versions of firmware~\cite{cui2013firmware}.
Moreover, the symbol information such as function names is generally stripped during the firmware compilation.
Finding such vulnerable functions simply based on function names is impossible. 
To this end, the binary code similarity detection (BCSD) technique is applied to quickly find such homologous vulnerabilities in a large amount of firmware~\cite{david2018firmup}.
The BCSD technique focuses on determining the similarity between two binary code pieces.
As to the vulnerability search, the BCSD focuses on finding other homologous vulnerable functions given a known vulnerability function.
In addition to the vulnerability search, BCSD has been widely used for other security applications such as code plagiarism detection~\cite{Basit:2005, schulman2005finding, luo2014semantics}, malware detection~\cite{hu2009large,hu2013mutantx}, and patch analysis~\cite{gao2008binhunt, wang2000bmat, dullien2005graph}. 
Despite many existing research efforts, the diversity of IoT hardware architectures and software platforms poses challenges to BCSD for IoT firmware. 
There are many different instruction set architectures (ISA) such as ARM, PowerPC, MIPS, and X86 for IoT firmware.
The instructions are different and the rules, such as the calling convention and the stack layout, also differ across different ISAs \cite{pewny2015cross}.
It is non-trivial to find homologous vulnerable functions across platforms.


BCSD can be generally classified into two categories: i) dynamic analysis-based method{s} and ii) static analysis-based method{s}.
The methods based on dynamic analysis capture the runtime behavior as function features by running a program, where the function features can be I/O pairs of function~\cite{pewny2015cross} or system calls during the program execution \cite{egele2014blanket}, etc.
But this kind of method is not scalable for large-scale firmware analysis since running firmware requires specific devices and emulating firmware is also difficult~\cite{zaddach2014avatar, gustafson2019toward, chen2016towards}.
The methods based on static analysis mainly extract function features from assembly code.
An intuitive way is to calculate the edit distance between assembly code sequences~\cite{David:Tracelet}.
But this method cannot be directly applied in cross-architecture BCSD since instructions are different across architectures.
Architecture-independent statistical features of functions are proposed for the similarity detection~\cite{eschweiler2016discovre}.
These features are less affected across architectures such as the number of function calls, strings, and constants.
Furthermore, the control flow graph (CFG) at the assembly code level is utilized by conducting a graph isomorphism comparison for improving the similarity detection ~\cite{eschweiler2016discovre, feng2016scalable}.
Based on statistical features and CFG, Gemini~\cite{xu2017neural} leverages the graph embedding network to encode functions to  vectors for similarity detection.
Static analysis-based methods are faster and more scalable for large-scale firmware analysis but often produce false positives due to the lack of semantic information.
Since homologous vulnerable functions in different architectures usually share the same semantics, it is desirable that a cross-platform BCSD can capture the function semantic information in a scalable manner.   

Abstract syntax tree (AST) is a tree representation of code, which is usually used in semantic analysis of source code compilation~\cite{jones2003abstract}, and preserves well-defined statement components, explicit order of statements, and the execution logic.
AST contains rich semantic information and is easy to generate by decompiling the function during binary analysis. 
Therefore, the semantics contained in an AST can be potentially used for identifying homologous functions across architectures.
Considering the similarities between the natural language and the assembly language such as words and instructions, sentences and basic blocks, and the successful application of natural language processing (NLP) technologies in programming language analysis~\cite{raff2018malware}, in this work, we propose \ProjectName{}, a deep learning-based AST-encoding method for cross-platform function similarity detection.

Tree-LSTM network has been shown to provide a better performance in predicting the semantic relatedness of two sentences than the plain LSTM~\cite{tai2015improved, treelstm}. 
We utilize the Tree-LSTM network to encode ASTs into semantic representation vectors. 
Given a binary program, we first extract ASTs by decompiling its functions. 
Then we preprocess the ASTs and encode them into representation vectors by Tree-LSTM for semantic representation.
After the encoding, we adopt the \siamese{}~\cite{chopra2005learning} to integrate two Tree-LSTM networks for calculating the similarity between the encoding vectors as the AST similarity.
Then, we further calibrate the AST similarity with function call relationships to calculate the final function similarity.
We construct a large number of cross-architecture AST pairs to train the Tree-LSTM network so that it can recognize the semantic equivalent AST pairs.
We conduct a series of evaluations, and the results show that our method outperforms the baseline approaches.
In terms of the similarity detection accuracy, our method \ProjectName{} outperforms \diaphora{}~\cite{diaphora} by $77\%$ and \gemini{} by $4.4\%$.
Remarkably, our method takes an average of {$8 \times 10^{-9}$ seconds for the similarity calculation of a pair of AST encodings}, which is $10^4$ times faster than \gemini{} and $10^6$ times faster than \diaphora{}.

\del{We complete a prototype of \ProjectName{}, and with it, we conduct a vulnerability search in a large number of firmware and find out 75 homologous vulnerable functions.
After statistical analysis of search results, We find that the same vulnerabilities such as CVE-2016-2105 exist in different device firmware and different versions of the same device firmware.
By comparing the functions in the results, we found that the higher the repetition of the code composition, the more similar the function semantics, the higher the similarity score.}

\noindent\textbf{Contribution.} The main contributions of this paper are as follows:\vspace{-2pt}
\begin{itemize} 
\item We present a new method named \ProjectName{}, to encode ASTs at binary level with the Tree-LSTM network into semantic representation vectors for BCSD.
   {To facilitate the model training, we utilize the \siamese{}  to integrate two identical Tree-LSTM model for AST similarity detection.}
   Furthermore, we introduce an additional similarity calibration scheme to improve the similarity calculation accuracy.

\item We implement and open-source a prototype of \ProjectName{}\footnotemark[1]. For the model training, we build a large-scale cross-platform dataset containing 49,725 binary files by cross-compiling 260 open-source software. 
We compare our model against the state-of-the-art approach \gemini{}~\cite{xu2017neural} and an AST-based method \diaphora{}~\cite{diaphora}.  
The evaluation results show that our \ProjectName{} outperforms \diaphora{} and \gemini{}.


\item 
In the application of IoT vulnerability search, we collect 5,979 firmware images and 7 vulnerable functions from CVE database~\cite{CVECVE38:online}. 
we perform a similarity detection between firmware functions and vulnerable functions.
\ProjectName{} successfully identified 75 vulnerable functions in IoT firmware images.
From the analysis of the search results, our method can efficiently and accurately capture the semantic equivalence of code fragments for the cross-platform BCSD. 

\end{itemize}   
\footnotetext[1]{\ProjectUrl{}}

\section{Preliminary}

In this section, we first briefly describe the AST structure.
Then we compare the architectural stability of AST and CFG across different platforms.
Next we introduce the Tree-LSTM network adopted in our method.
Finally, we introduce the problem definition of function similarity calculation.

\begin{figure}[thb]
    \centering
    \includegraphics[scale=0.4]{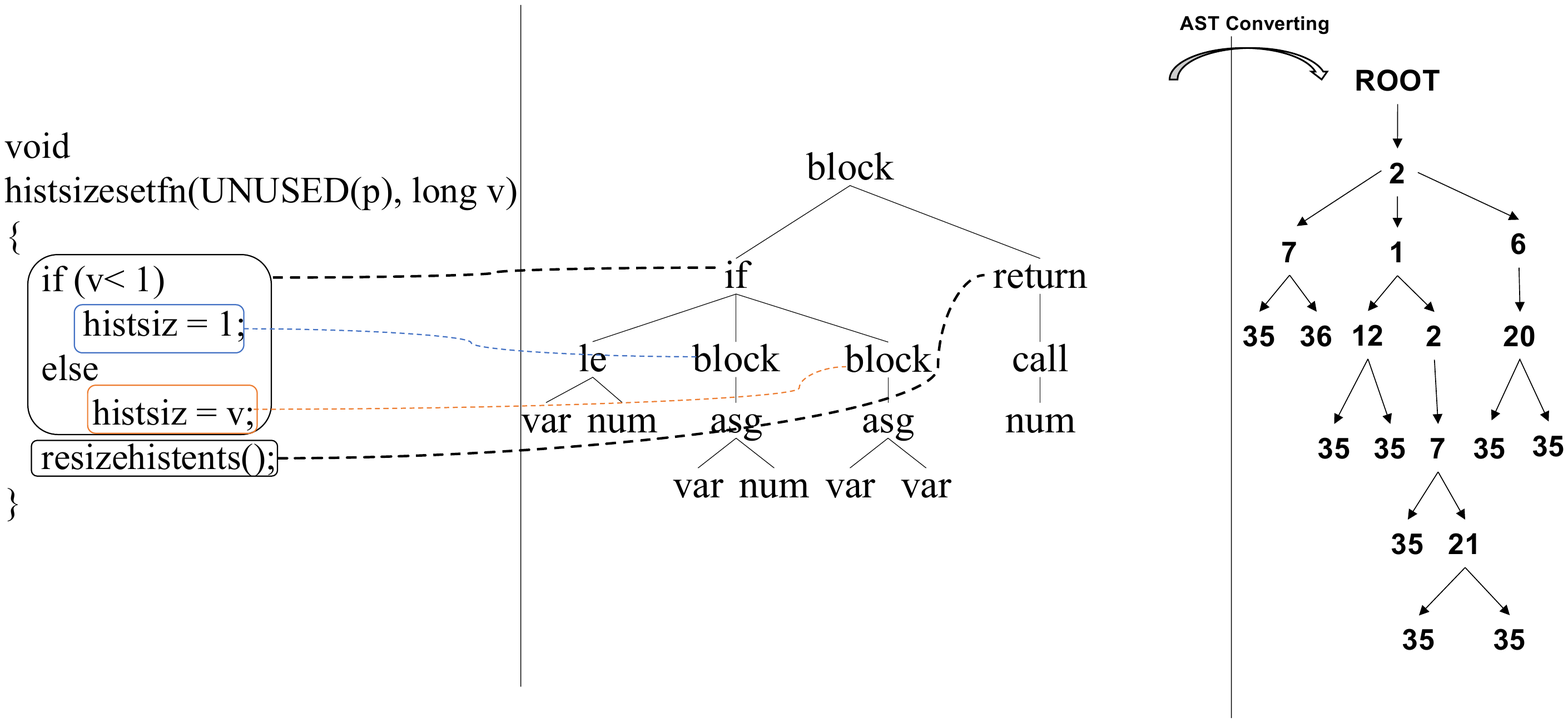}\vspace{-6pt}
    \caption{Source code of function \textit{histsizesetfn} and the corresponding decompiled AST of x86 architecture. }
    \label{fig:source_to_ast}\vspace{-10pt}
\end{figure}

\subsection{Abstract Syntax Tree} \label{sec:ast_intro}
An AST is a tree representation of the abstract syntactic structure of code in the compilation and decompilation process.
Different subtrees in an AST correspond to different code scopes in the source code.
Figure \ref{fig:source_to_ast} shows a decompiled AST corresponding to the source code of function \texttt{histsizesetfn} in \texttt{zsh v5.6.2} in the left. 
The \texttt{zsh} is a popular shell software designed for interactive use, and the function \texttt{histsizesetfn} sets the value of a parameter.
The lines connecting the source code and AST in Figure~\ref{fig:source_to_ast} show that {a node} in the AST corresponds to an expression or a statement in the source code.
A variable or a constant value is represented by a leaf node in AST.
We group nodes in an AST into two categories: i) statement nodes and ii) expression nodes according to their functionalities shown in Table~\ref{tab:mapping}.
Statement nodes control the function execution flow while expression nodes perform various calculations.
Statement nodes include \nodename{if, for, while, return, break} and so on.
Expression nodes include common arithmetic operations and bit operations.

\begin{table}[!h]
    \centering
    \resizebox{8cm}{!}{
    \begin{tabularx}{0.5\textwidth}{|c|c|c|X|}
        \hline
         & \cellcolor[HTML]{C0C0C0}  Node Type & \cellcolor[HTML]{C0C0C0} Label & \cellcolor[HTML]{C0C0C0} Note  \\ 
         \hline
         \multirow{10}*{Statement} 
         ~& if   & 1 & if statement   \\
         ~& block   & 2& instructions executed sequentially\\
         ~& for   & 3 & for loop statement\\
         ~& while   & 4 & while loop statement\\
         ~& switch   & 5 & switch statement\\
         ~& return   & 6 &return statement \\
         ~& goto   & 7 & unconditional jump \\
         ~& continue   & 8 &continue statement in a loop\\
         ~& break   & 9 &break statement in a loop \\
         \hline
         \multirow{15}*{Expression}
            & asgs   & 10$\sim$17 & assignments, including assignment, assignment after or, xor, and, add, sub, mul, div\\
             & cmps & 18$\sim$23  & comparisons including equal, not equal, greater than, less than, greater than or equal to, and less than or equal to.\\
             & ariths & 24$\sim$34& arithmetic operations including or, xor, addition, subtraction, multiplication, division, not, post-increase, post-decrease, pre-increase, and pre-decrease \\
             & other & 34$\sim$43 & others including indexing, variable, number, function call, string, asm, and so on.\\
         \hline 
    \end{tabularx}
    }
    \caption{Statements and Expressions in ASTs. We count the statements and expressions for nodes in ASTs after the decompilation by \texttt{IDA Pro} and list the common statements and expressions. This table can be extended if new statements or expressions are introduced.}
    \label{tab:mapping}	\vspace{-10pt}
\end{table}

\begin{figure}[tbh]
\centering
    \begin{minipage}[t]{0.23\textwidth}
    \centering
    \includegraphics[scale=0.35]{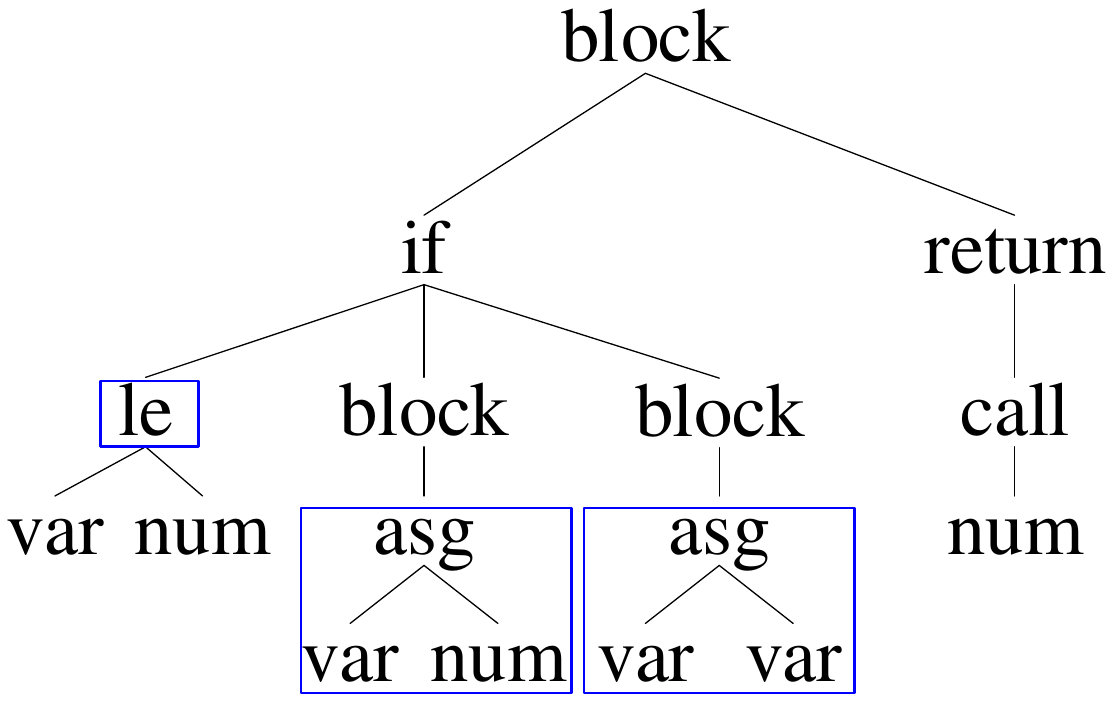}
    \end{minipage}
    \begin{minipage}[t]{0.23\textwidth}
    \centering
    \includegraphics[scale=0.35]{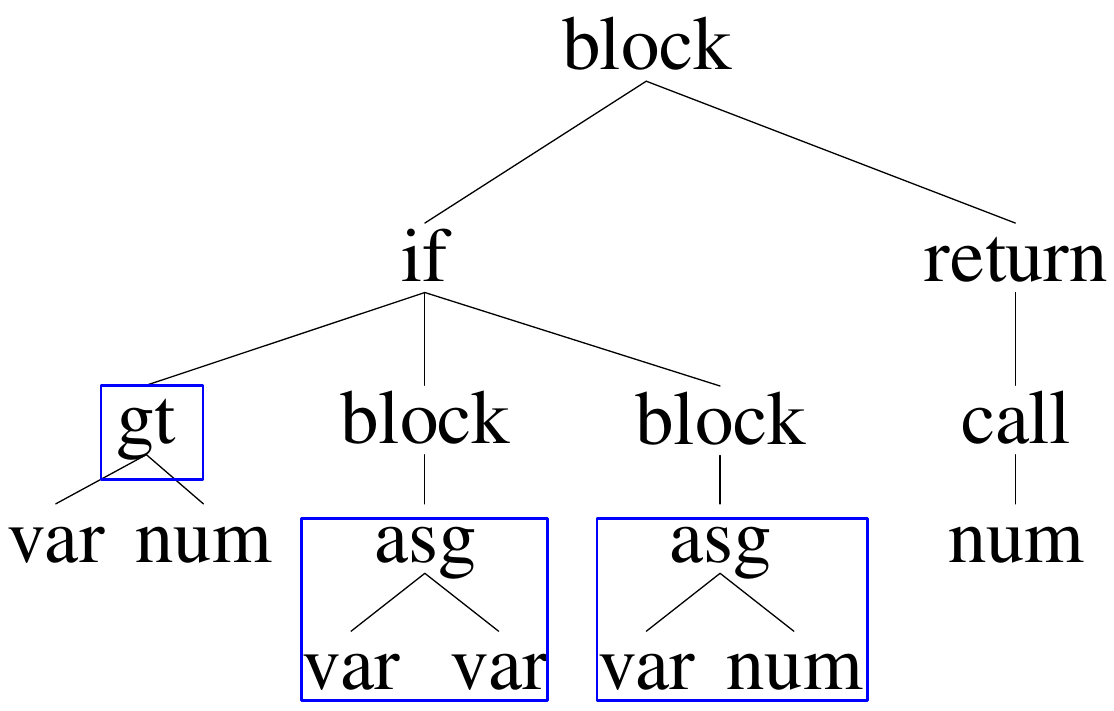}
    \end{minipage}
    \begin{minipage}[t]{0.23\textwidth}
    \vspace{-12pt}
    \subcaption{AST for x86 platform}
    \end{minipage}
    \begin{minipage}[t]{0.23\textwidth}
    \vspace{-12pt}
    \subcaption{AST for ARM platform}
    \end{minipage}
    \begin{minipage}[t]{0.23\textwidth}
    \centering
    \includegraphics[scale=0.29]{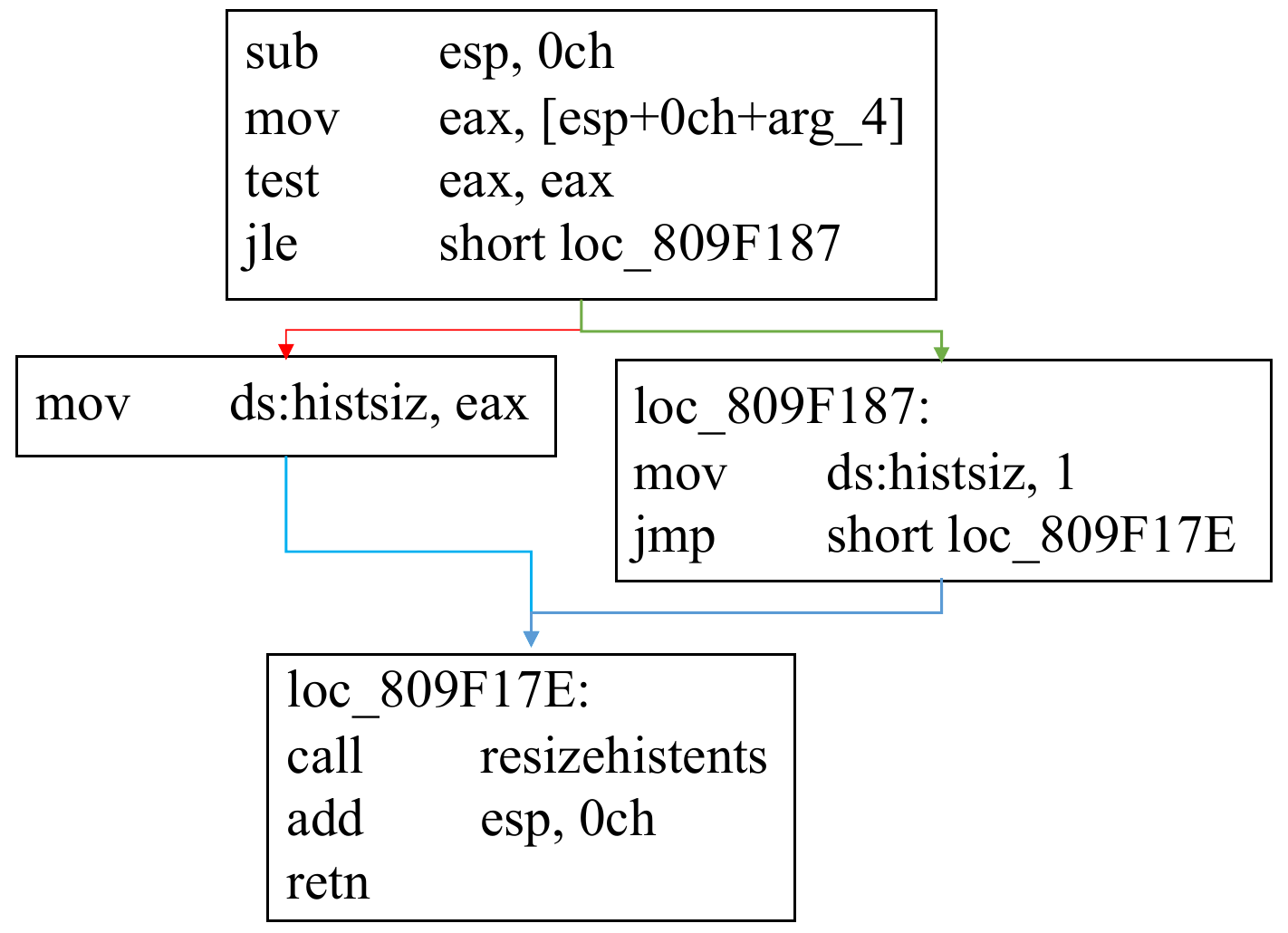}
    \end{minipage}
     \begin{minipage}[t]{0.23\textwidth}
    \centering
    \includegraphics[scale=0.35]{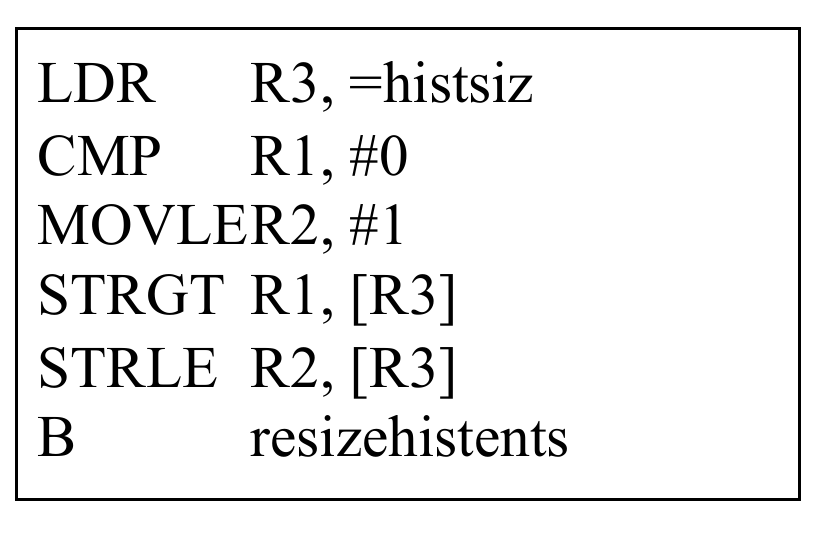}
    \end{minipage}
    \begin{minipage}[t]{0.23\textwidth}
     \vspace{-12pt}
    \subcaption{CFG for x86 platform}
    \end{minipage}
    \hfill
    \begin{minipage}[t]{0.23\textwidth}
    \vspace{-12pt}
    \subcaption{CFG for ARM platform}
    \end{minipage}\vspace{-8pt}    
    \caption{ASTs and CFGs of the function \textit{histsizesetfn} under different architectures.}
    \label{fig:comparisonofcfg}\vspace{-15pt}
\end{figure}

\begin{figure*}[bht]
    \centering
    \includegraphics[scale=0.6]{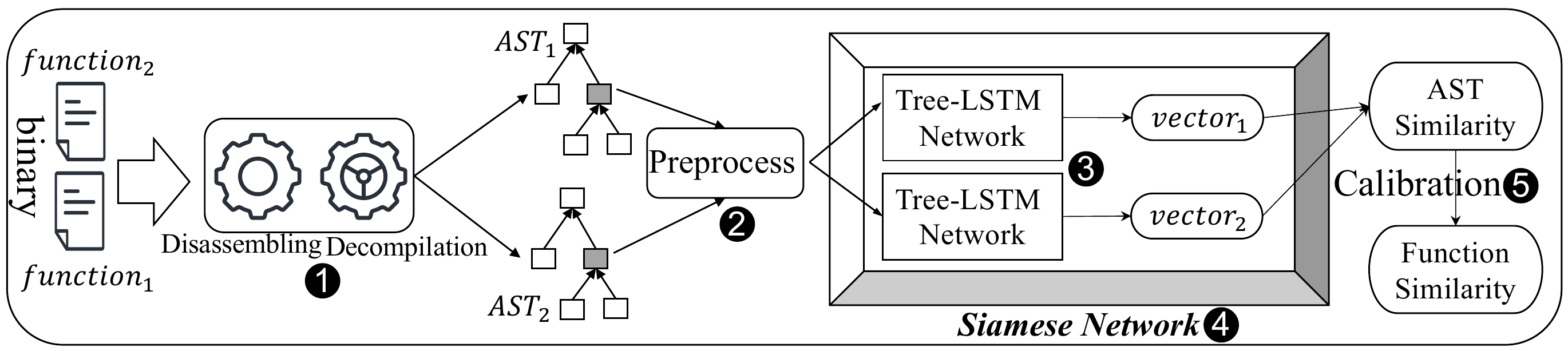}
    \vspace{-6pt}
    \caption{System Overview. The workflow of the system is composed of five steps: AST extraction ({\large{\ding{202}}}), AST preprocessing ({\large{\ding{203}}}), AST encoding ({\large{\ding{204}}}), similarity calculation by \siamese{} ({\large{\ding{205}}}), and similarity calibration ({\large{\ding{206}}}).}
    \label{fig:overview}\vspace{-10pt}
\end{figure*}

\subsection{AST \vs{} CFG} \label{sec:astvscfg}
Both CFG and AST are structural representations of a function.
The CFG of a function contains the jump relationships between basic blocks that contain straight-line code sequences~\cite{hennessy2011computer}.
Though CFG has been used for similarity measurement in BCSD~\cite{eschweiler2016discovre}, David {\etal}~\cite{David:Tracelet} demonstrated that CFG structures are greatly affected by different architectures. We find AST shows  better architectural stability across architectures compared with CFG \clnote{since the AST is generated from the machine independent intermediate representations which are disassembled from assemble instructions during the decompilation process~\cite{cifuentes1995decompilation}.} 
Figure~\ref{fig:comparisonofcfg} shows the changes of ASTs and CFGs in x86 and ARM architectures, respectively. For the CFGs from x86 to ARM, we observe that the number of basic blocks changes from 4 to 1, and the number of assembly instructions has changed a lot.
However, the ASTs, \clnote{based on higher level intermediate representation,} slightly change from x86 to ARM, where the changes are highlighted with blue boxes.
Besides, AST preserves the semantics of functionality and is thus an ideal structure for cross-platform similarity detection.




\subsection{Tree-LSTM}
In natural language processing, Recursive Neural Networks (RNN) are widely applied and perform better than Convolutional Neural Networks~\cite{yin2017comparative}.
RNNs take sequences of arbitrary lengths as inputs considering that a sentence can consist of any number of words.
However, standard RNNs are not capable of handling long-term dependencies due to the gradient vanishing and gradient exploding problems.
As one of the variants of RNN, Long Short-Term Memory (LSTM)~\cite{HochreiterLong} has been proposed to solve such problems. 
LSTM introduces a gate mechanism including the input, forget, and output gates. As described in \cref{sec:modeldesc}, the gates control the information transfer to avoid the gradient vanishing and exploding.
Nevertheless, LSTM can only process sequence input but not structured input.
Tree-LSTM is proposed to process tree-structured inputs~\cite{tai2015improved}.
The calculation by Tree-LSTM model is from the bottom up.
For each non-leaf node in the tree, all information from child nodes is gathered and used for the calculation of the current node.
In sentiment classification and semantic relatedness tasks, Tree-LSTM performs better than a plain LSTM structure network.
There are two types of Tree-LSTM proposed in the work \cite{treelstm}: Child-Sum Tree-LSTM and Binary Tree-LSTM.
Researchers have shown that Binary Tree-LSTM performs better than Child-Sum Tree-LSTM~\cite{treelstm}.
Since the Child-Sum Tree-LSTM does not take into account the order of child nodes, while the order of statements in AST reflects the function semantics, we use the Binary Tree-LSTM for our AST encoding.

\subsection{Problem Definition}
Given two binary functions ${F}_1$ and ${F}_2$, let
${T}_1$ and ${T}_2$ denote the corresponding ASTs of ${F}_1$ and ${F}_2$ which can be extracted after the decompilation of binary code. An AST is denoted as ${T=<V,E>}$ where ${V}$ and ${E}$ are the sets of vertices and edges, respectively. In node set ${V} = \{v_1, v_2, ..., v_k, ..., v_n\}$, {every node corresponds to a number listed in Table~\ref{tab:mapping}, and $n$ is the number of vertices in the AST. 
For an edge $l_{kj} $ in ${E}$, it means vertex $v_k$ and $v_j$ are connected, and $v_k$ is the parent node of $v_j$. 
Given two ASTs ${T}_1$ and ${T}_2$, we define a model $\mathcal{M}({T}_1, {T}_2)$ to calculate the similarity between them, where the similarity score ranges from 0 to 1. 
In an ideal case, when ${F}_1$ and ${F}_2$ are homologous, the model $\mathcal{M}({T}_1, {T}_2)$ is expected to output a score of 1 (in an ideal case). And when ${F}_1$ and ${F}_2$ are non-homologous, $\mathcal{M}({T}_1, {T}_2)$ is expected to output a score of 0.
In addition, based on the semantic similarity of ASTs, we leverage the numbers of callee functions of ${F}_1$ and ${F}_2$ for similarity calibration.
Let $C_1$ and $C_2$ denote the numbers of callee functions corresponding to functions $F_1$ and $F_2$, respectively. We define the calibration function $\mathcal{S}(C_1, C_2)$, where the range is also from 0 to 1.
The final function similarity calculation is defined as $\mathcal{F}({F_1}, {F_2}) = \mathcal{M}(T_1, T_2) \times \mathcal{S}(C_1, C_2)$.
The larger the score, the higher similarity for the two binary functions ${F}_1$ and ${F}_2$.

\section{Methodology}

\label{sec:methodology}

In this section, we first present the overview of \ProjectName{}.
Then we detail the similarity calculation for two ASTs, consisting of the AST encoding of {$\mathcal{N}(T)$} and the similarity calculation between the encodings of $\mathcal{M}({T}_1, {T}_2)$. 
Then we introduce the calibration in the final function similarity calculation of $\mathcal{F}({F_1}, {F_2})$.

\subsection{Overview}\label{sec:overview}
Figure \ref{fig:overview} delineates the workflow of \ProjectName{}.
Given two binary functions, we first disassemble and decompile them to extract the ASTs.
Before encoding the ASTs into the Tree-LSTM network, AST preprocessing is performed, including the digitalization and format transformation.
AST digitalization means that each node in an AST is mapped and replaced with an integer value, where 
the specific mapping relationship is shown in Table~\ref{tab:mapping}.
For example, the node of an assignment statement is replaced with a value of 10.
Since we adopt the Binary Tree-LSTM~\cite{tai2015improved}, which takes binary trees as inputs, ASTs are transformed to the \textit{left-child right-sibling} format \cite{ruskey1977generating} after the digitalization.
After that, each node's first child node in the original AST is made its left child and its nearest sibling to the right in the original tree is made its right child.
Next, the two ASTs are encoded into two vectors by the two Tree-LSTM networks in \siamese{}~\cite{chopra2005learning}.
Then the \siamese{} utilizes two encoding vectors to compute the similarity as the AST similarity (\cref{sec:modeldesc}).
Finally, we use the function calling relationship to calibrate the AST similarity (\cref{sec:simcor}). 

\begin{figure*}[htb]
\centering
\includegraphics[scale=0.55]{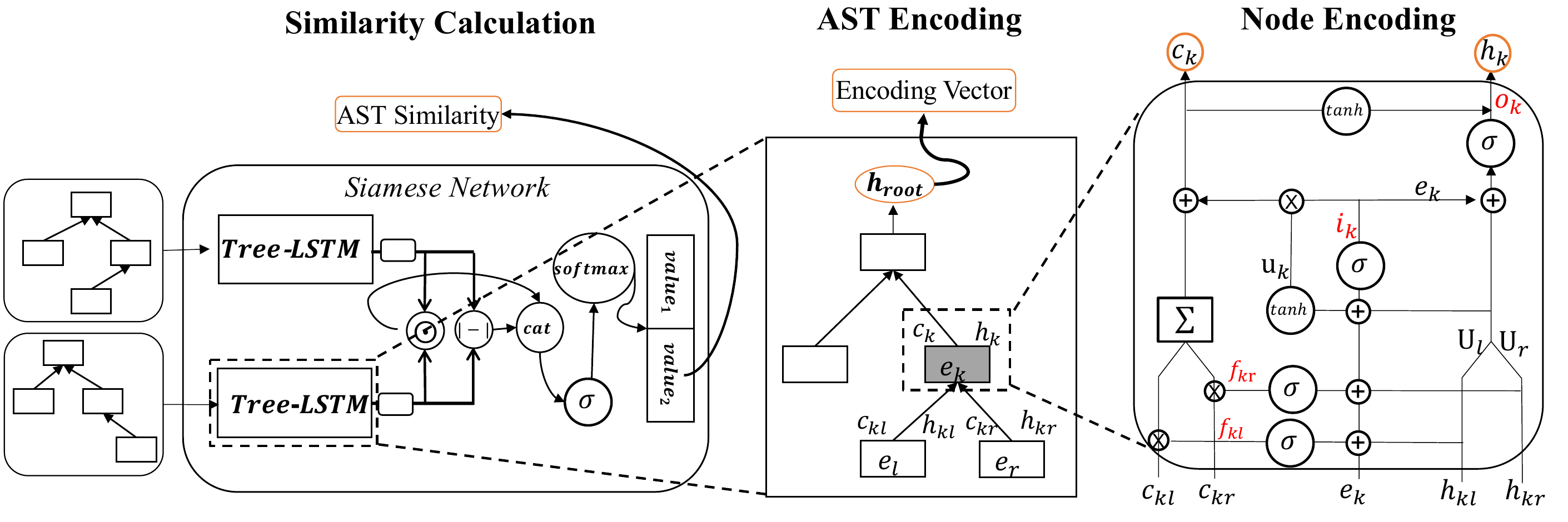}
\caption{Different views of the similarity calculation. The workflow of similarity calculation of two ASTs in \siamese{}  is shown in the left. The AST encoding, as part of \siamese{}, is shown in the middle. The node encoding by Tree-LSTM, as part of AST encoding, is shown in the right.}
\label{fig:calculation}
\vspace{-15pt} 
\end{figure*}
\subsection{AST Similarity Calculation}\label{sec:modeldesc}
We present the Tree-LSTM network and \siamese{}. 
The Tree-LSTM network $\mathcal{N}(\cdot)$ is used for AST encoding.
The \siamese{} utilizes the outputs of the Tree-LSTM network for calculating the AST similarity.
\paragraph*{\textbf{AST Encoding}}\label{sec:encoding}
Tree-LSTM network is firstly proposed to encode the tree representation of a sentence and summarize the semantic information in natural language processing.
Tree-LSTM network can preserve every property of the plain LSTM gating mechanisms while processing tree-structured inputs.
The main difference between the plain LSTM and the Tree-LSTM is the way to deal with the outputs of predecessors.
The plain LSTM utilizes the output of only one predecessor in the sequence input.
We utilize Tree-LSTM to integrate the outputs of all child nodes in the AST for calculation of the current node. 
The process of node encoding for node $v_k$ in Tree-LSTM is shown in Figure \ref{fig:calculation}.
The Tree-LSTM takes three types of inputs: node embedding $e_{k}$, hidden states $h_{kl}$ and $h_{kr}$, and cell states $c_{kl}$ and $c_{kr}$.
The node embedding $e_{k}$ is generated by embedding the node $v_k$ to a high-dimensional representation vector.
$h_{kl}$, $h_{kr}$, $c_{kl}$, and $c_{kr}$ are outputs from the encoding of child nodes.
During the node encoding in Tree-LSTM, there are three gates and three states which are important in the calculation.
The three gates are calculated for filtering information to avoid gradient explosion and gradient vanishing~\cite{tai2015improved}.
They are input, output, and forget gates.
In Binary Tree-LSTM, there are two forget gates $f_{kl}$ and $f_{kr}$, filtering the cell states from the left child node and right child node separately.
As shown in \textbf{Node Encoding} in Figure \ref{fig:calculation}, the forget gates are calculated by combining $h_{kl}$, $h_{kr}$, and $e_k$.
Similar to the forget gates, the input gate, and the output gate are also calculated by combining $h_{kl}$, $h_{kr}$, and $e_k$.
The details of the three types of gates are as follows:
\begin{equation}
\label{equation:fl}
\begin{aligned}
   f_{kl} = \sigma(W^{f}e_k+(U^{f}_{ll}h_{kl}+U^{f}_{lr}h_{kr})+b^{f})
\end{aligned}
\end{equation}
\begin{equation}
\label{equation:fr}
\begin{aligned}
   f_{kr} = \sigma(W^{f}e_k+(U^{f}_{rl}h_{kl}+U^{f}_{rr}h_{kr})+b^{f})
\end{aligned}
\end{equation}
\begin{equation}
\label{equation:ik}
\begin{aligned}
   i_{k} = \sigma(W^{i}e_k+(U^{i}_lh_{kl}+U^{i}_rh_{kr})+b^{i})
\end{aligned}
\end{equation}
\begin{equation}
\label{equation:ok}
\begin{aligned}
   o_{k} = \sigma(W^{o}e_k+(U^{o}_{l}h_{kl}+U^{o}_{r}h_{kr})+b^{o})
\end{aligned}
\end{equation}
where $i_k$ and $o_k$ denote the input gate and the output gate respectively,
and the symbol $\sigma$ denotes the \texttt{sigmoid} activation function.
The weight matrix $W$, $U$, and bias $b$ are different corresponding to different gates.
After the gates are calculated, there are three states $u_k$, $c_k$,  and $h_k$ in Tree-LSTM to store the intermediate encodings calculated based on inputs $h_{kl}$, $h_{kr}$, and $e_k$.
The cached state $u_k$ combines the information from the node embedding $e_k$ and the hidden states $h_{kl}$ and $h_{kr}$ {(Equation \ref{equation:uk}).}
And note that $u_k$ utilizes \texttt{tanh} as the activation function rather than $sigmoid$ for holding more information from the inputs.
The cell state $c_k$ combines the information from the cached state $u_k$ and the cell states $c_{kl}$ and $c_{kr}$ filtered by forget gates (Equation \ref{equation:ck}).
The hidden state $h_k$ is calculated by {combining the information} from cell state $c_k$ and the output gate $o_k$ (Equation \ref{equation:hk}).
The three states are computed as follows:
\begin{equation}
\label{equation:uk}
\begin{aligned}
   u_{k} = tanh(W^{u}e_k+(U^{u}_{l}h_{kl}+U^{u}_{r}h_{kr})+b^{u})
\end{aligned}
\end{equation}
\begin{equation}
\label{equation:ck}
\begin{aligned}
   c_{k} =  i_k \odot u_k +( c_{kl} \odot f_{kl} + c_{kr} \odot f_{kr})
\end{aligned}
\end{equation}
\begin{equation}
\label{equation:hk}
\begin{aligned}
   h_k = o_k \odot tanh(c_k)
\end{aligned}
\end{equation}
where the $\odot$ means Hadamard product~\cite{horn1990hadamard}.
After the hidden state and input state are calculated, the encoding of the current node $v_k$ is finished.
The states $c_k$ and $h_k$ will then be used for the encoding of $v_k$'s parent node.
During the AST encoding, Tree-LSTM encodes every node in the AST from bottom up as shown in \textbf{AST Encoding} in Figure~\ref{fig:calculation}.
After encoding all nodes in the AST, the hidden state of the root node is used as the encoding of the AST.

\paragraph*{\textbf{\siamese{}}}
After two ASTs are encoded by Tree-LSTM, the \siamese{} utilizes AST encodings to calculate the similarity between these two ASTs.
The details of the \siamese{} $\mathcal{M}({T}_1, {T}_2)$ are shown in Figure \ref{fig:calculation}.
The \siamese{} consists of two identical Tree-LSTM networks that share the same parameters.
In the process of similarity calculation, the \siamese{} first utilizes Tree-LSTM to encode ASTs into vectors.
We design the \siamese{} with subtraction and multiplication operations to capture the relationship between the two encoding vectors.
After the operations, the two resulting vectors are concatenated into a larger vector.
Then the resulting vector goes through a layer of \textit{softmax} function to generate a 2-dimensional vector.
The calculation is defined as:
\begin{equation}
\label{equation:classification}
\begin{aligned}
   &\mathcal{M}({T}_1, {T}_2) =\\
  &softmax(\sigma( cat(|\mathcal{N}({T}_1) - \mathcal{N}({T}_2)|, \mathcal{N}({T}_1) \odot \mathcal{N}({T}_2)) \times W )))
\end{aligned}
\end{equation}
where $W$ is a $2n \times 2$ matrix, the $\odot$ represents Hadamard product~\cite{horn1990hadamard}, $|\cdot|$ denotes the operation of making an absolute value, the function $cat(\cdot)$ denotes the operation of concatenating vectors.
The softmax function normalizes the vector into a probability distribution.
Since $W$ is a $2n \times 2$ weight matrix, the output of \siamese{} is a $2\times 1$ vector.
The format of output is $[\small{dissimilarity\text{ }score},\small{similarity\text{ }score}]$, where the first value represents the dissimilarity score and the second represents the similarity score.
During the model training, the input format of \siamese{} is $<{T}_1, {T}_2, label>$. 
In our work, the label vector $[1,0]$ means ${T}_1$ and ${T}_2$ are from non-homologous function pairs and the vector $[0,1]$ means homologous. 
The resulting vector and the label vector are used for model loss and gradient calculation.
During model inference, the second value in the output vector is taken as the similarity of the two ASTs.

\begin{figure}[h]
    \centering
    \includegraphics[scale=0.49]{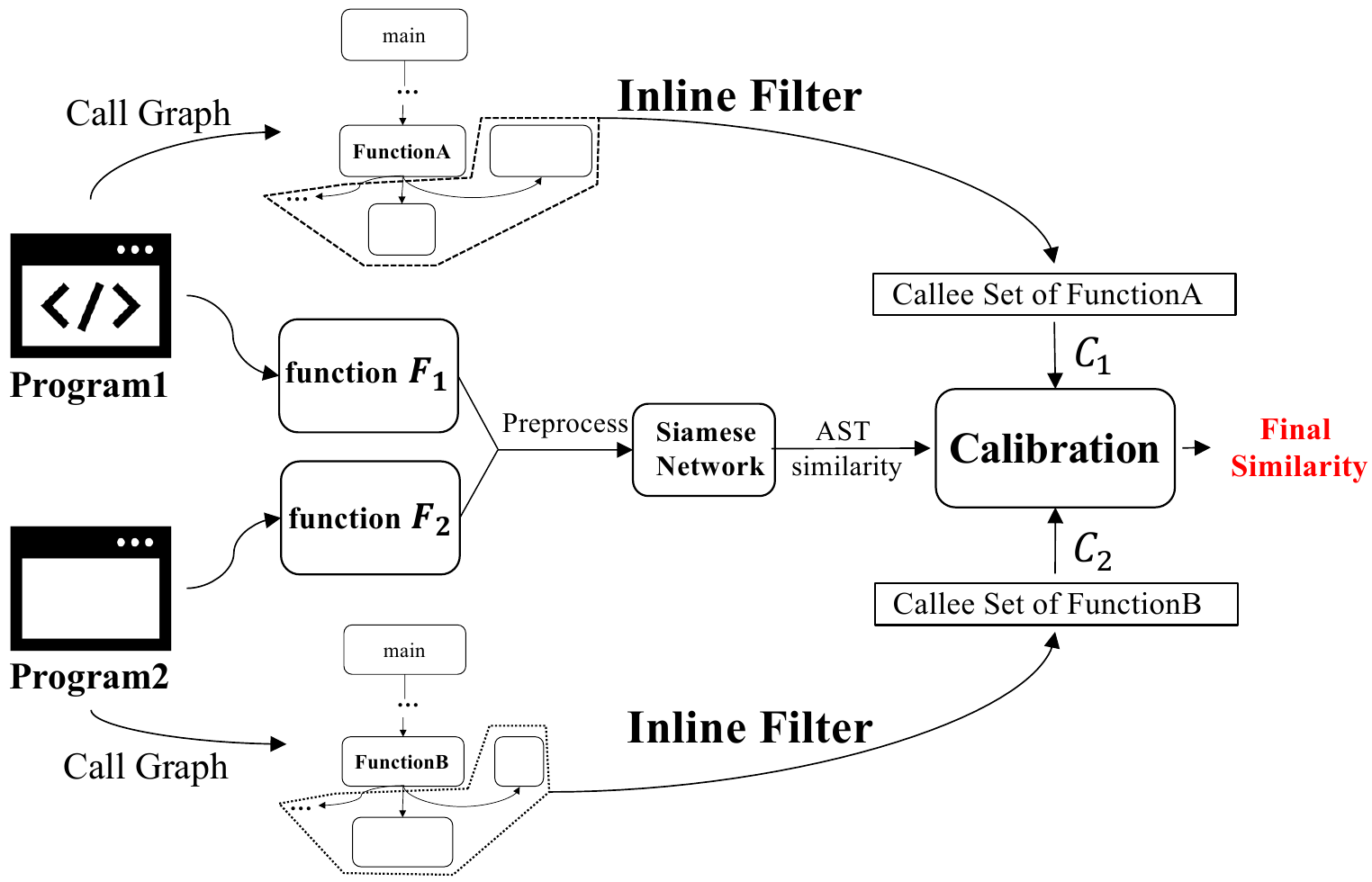}
    \caption{\clnote{Overview of AST Similarity Calibration.}}
    \label{fig:calibrationexample}
    \vspace{-10pt}
\end{figure}
\subsection{\textbf{AST Similarity  Calibration}}\label{sec:simcor}
Considering other potential attributes (e.g., the number of variables and statements) change with different architectures as shown in Figure~\ref{fig:comparisonofcfg} (c) and (d), the number of callee functions is an architecture-independent feature. The number of calllee functions is easy to count during the reverse engineering.
We combine the number of the callee functions and AST similarity to calculate the function similarity as shown in Figure~\ref{fig:calibrationexample}.
The callee functions of function $F$ are the functions called in function $F$.
Homologous functions likely have the same number of callee functions since they share the same \clnote{source code}.
\clnote{Considering that function inlining may occur during compilation~\cite{ayers1997aggressive,zhao2003inline}, which will reduce the number of callee functions, we refine callee functions by filtering the functions that may be inlined.
Specifically, we filter out the callee functions whose number of assembly instructions are less than the threshold $\beta{}$, and the rest are used as the callee function set $\chi{}$ of function F.}
With the introduction of the callee functions, non-homologous functions which may generate similar ASTs can be better distinguished.
We define $C$ as the size of the callee function set $\chi{}$ of function $F$.
We use the size $C$ to calibrate the AST similarity.
We use the exponential function to measure the similarity between callee functions.
The calculation for callee functions is as follows:
\begin{equation}
\label{equation:sigmoid}
\begin{aligned}
   \mathcal{S}({C_1}, {C_2}) = e^{-|C_1 - C_2|}
\end{aligned}
\end{equation}
where $C_1$ and $C_2$ are the size of callee function set of binary function $F_1$ and $F_2$, respectively.
The final function pair similarity calculation combines the AST similarity and the callee function similarity is as follows:
\begin{equation}
\label{equation:sigmoid}
\begin{aligned}
   \mathcal{F}({F_1}, {F_2}) = \mathcal{M}(T_1, T_2) \times \mathcal{S}({C_1}, {C_2})
\end{aligned}
\end{equation}
where $T_1$ is the AST decompiled from binary function $F_1$ and $T_2$ is the AST decompiled from binary function $F_2$.
We conduct a comparative evaluation in \cref{sec:roc} to show the performance gain of our AST similarity calibration. 

\section{Evaluation}

\subsection{Experiment Settings}\label{sec:modelsetting}
We utilize \texttt{IDA Pro 7.3} \cite{ida} and its plug-in \textit{Hexray Decompiler} to decompile binary code for AST extraction. 
Since the \textit{Hexray Decompiler} currently only supports the architectures of x86, x64, PowerPC (PPC), and ARM, our approach cannot handle the binaries which belong to the MIPS architecture. 
After the AST extraction, we perform preprocessing of AST as described in \cref{sec:overview}.
Before the AST encoding by the Tree-LSTM, the nodes in an AST are embedded into 16-dimensional vectors by the \textit{nn.Embedding} of PyTorch~\cite{url:pytorch}.
For the encoding of leaf nodes in Formulas (\ref{equation:fl})-(\ref{equation:ck}), we assign the state vectors $h_{kl}$, $h_{kr}$, $c_{kl}$, and $c_{kr}$ to zero vectors.
The loss function for model training is \textit{BCELoss}, which measures the binary cross entropy between the labels and the predictions. 
The \textit{AdaGrad} optimizer is applied for gradient computation and weight-matrix updating after losses are computed. 
Since the The computation steps of Tree-LSTM depend on the shape of the AST, therefore, we cannot perform parallel batch computation, which makes the batch size always to be 1. 
The model is trained for 60 epochs.
We do not include the calibration scheme introduced in \cref{sec:simcor} during the model training, so that the Tree-LSTM network effectively learns semantic differences between ASTs.
Our experiments are performed on a local server equipped with two Intel(R) Xeon(R) CPUs E5-2620 v4 @ 2.10GHz, each with 16 cores, 128GB of RAM, and 4T of storage. The code of \ProjectName{} runs in a Python 3.6 environment. We use \texttt{gcc v5.4.0} compiler to compile source code in our dataset, and use the \texttt{buildroot-2018.11.1}~\cite{url:buildroot} for the Buildroot dataset (details in \cref{dataset}) construction. We use the tool binwalk~\cite{url:binwalk} to unpack the firmware for obtaining the binaries to conduct further analysis.

\begin{table}[!h]
	\vspace{-6pt}
	\centering
	\resizebox{6.5cm}{!}{
		{
			\begin{tabular}{ccrr} 
    \hline
\rowcolor[HTML]{C0C0C0}      {{Name}} &{{Platform}}&  {\# of binaries} & {\# of functions} \\
     \hline
     \multirow{4}{*}{OpenSSL} & ARM & 6  &6,401  \\
     \cline{2-4}
      & x86 & 6  &6,420  \\
     \cline{2-4}
     & x64 & 6 & 6,539  \\
     \cline{2-4}
     & PPC & 6 & 6,317  \\
     \hline
     \multirow{4}{*}{Buildroot} & ARM & 10,142  &1,139,168  \\
     \cline{2-4}
      & x86 & 17,823  &1,765,547  \\
     \cline{2-4}
     & x64 & 11,005 & 1,088,411  \\
     \cline{2-4}
     & PPC & 10,755 & 1,133,647  \\
     \hline 
          \multirow{4}{*}{Firmware} & ARM & 5,661  &1,869,601 \\
     \cline{2-4}
      & x86 & 129  &70,400  \\
     \cline{2-4}
     & x64 & 202 & 79,383  \\
     \cline{2-4}
     & PPC & 1,098 & 392,268  \\
     \hline 
     \rowcolor[gray]{0.9} Total & & 56,839 & 7,564,102\\
     \hline
			\end{tabular} 
		}
	}
	\vspace{0pt}
	\caption{Number of binaries and functions in our datasets.}
    \label{tab:dataset}\vspace{-10pt}
\end{table}	

\subsection{Datasets}\label{dataset}
We create three datasets: \ysgbf{Buildroot dataset}, \ysgbf{OpenSSL dataset}, and \ysgbf{Firmware dataset}. 
The \ysgbf{Buildroot dataset} is used for model training and testing.
We obtain the optimal model weights with the best performance based on the results of model training and model testing (\cref{ModelSettings}). 
{The \ysgbf{OpenSSL dataset} is used for the comparative evaluation in \cref{sec:roc}}.
The \ysgbf{Firmware dataset} is used to perform the task of practical vulnerability search  with our approach.
As shown in Table~\ref{tab:dataset}, the third column shows the number of binary files for different architectures, and the fourth column shows the number of functions in the corresponding dataset.
Noting that in OpenSSL and Buildroot datasets, the symbols such as function names in binaries are retained during the compilation while these symbols are usually stripped in the release version of firmware.
We utilize the function names together with the library names since function names in different binaries might be duplicated but having different functionalities to construct function pairs with ground truth labels for the model training, testing, and evaluation.
Functions that have the same name from the same binary are called homologous function pairs, otherwise, they are considered non-homologous function pairs.
After the compilation, each function $F$ in the same source code corresponds to the different versions of binary functions \ysgbf{$F_{ARM}$, $F_{x86}$, $F_{x64}$,} and \ysgbf{$F_{PPC}$}.
The two binary functions which are homologous, for example \ysgbf{$F_{ARM}$} and \ysgbf{$F_{x86}$}, form the homologous pair \ysgbf{$(F_{ARM}, F_{x86})$}.
Two ASTs corresponding to the homologous pair are extracted and form the homologous AST pair \ysgbf{$(T_{x86}^F, T_{ARM}^F, +1)$}, where the ground truth label "$+1$" indicates that the two ASTs are from homologous functions.
The non-homologous functions, such as \ysgbf{$G_{x86}$} and \ysgbf{$F_{ARM}$}, form a non-homologous AST pair \ysgbf{$(T_{x86}^F, T_{ARM}^G, -1)$}, where "$-1$" indicates that the two ASTs are from non-homologous functions, and F and G are two different functions.
We construct six different architecture combinations of function pairs including ARM \vs{} PPC, ARM \vs{} x64, PPC \vs{} x64, x86 \vs{} ARM, x86 \vs{} PPC, and x86 \vs{} x64.
Function names in the Firmware dataset are deleted, \texttt{IDA} tags them with name \textit{sub\_xxx}, where \textit{xxx} is the function offset address.

\paragraph*{\textbf{Buildroot Dataset}} 
We build this dataset for model training and testing with a multi-platform cross-compilation tool \texttt{buildroot}~\cite{url:buildroot}.
We utilize the \texttt{buildroot} to download the source code of 260 different software which contains benchmark datasets including \texttt{busybox}, \texttt{binutils}, and compile them with the default compilation option for different architectures.
As a result, each software has four different binary versions corresponding to different architectures.
As shown in Table \ref{tab:dataset}, 49,725 binaries in total are compiled by the \texttt{buildroot}.
Then we randomly select functions from different architecture binaries to construct $1,022,616$ cross-architecture function pairs.
The details of function pairs of different architecture combinations are shown in Table \ref{tab:funcpair}.
Since we remove the AST pairs which contain ASTs with the node number less than 5, the number of function pairs are different among different architecture combinations.
We divide the Buildroot dataset into two parts according to the ratio of 8:2 for the training set and testing set.

\paragraph*{\textbf{OpenSSL Dataset}}
Considering Gemini uses OpenSSL dataset for the evaluation, to conduct a fair comparative experiments, we also build the OpenSSL dataset.
We compile the source code of \texttt{OpenSSL 1.1.0a}~\cite{url:openssl} under four different architectures with the default compilation settings.
We randomly select function pairs with different architecture combinations. 
The final dataset consists of 37,541 homologous AST pairs and 57,537 non-homologous AST pairs.

\paragraph*{\textbf{Firmware Dataset}}
We download 5,979 firmware from the websites and FTP servers of IoT device vendors. 
There are 2,300 firmwares from \texttt{NetGear}~\cite{url:netgear}, 1,021 from \texttt{Schneider}~\cite{url:schneider}, and 3,679 from \texttt{Dlink}~\cite{url:dlink}.
After unpacking firmware with \texttt{binwalk}~\cite{url:binwalk}, 7,090 binary files are generated. 
Note that not all firmware can be unpacked since \texttt{binwalk} cannot identify certain firmware format.
As shown in Table~\ref{tab:dataset}, binaries mainly come from the ARM and the PPC architectures.

\begin{table}	\centering
	\resizebox{7.0cm}{!}{
\begin{tabular}{cccc}
\hline
\rowcolor[HTML]{C0C0C0}  Arch-Comb&\# of pairs&Arch-Comb&\# of pairs\\
\hline
x86-ARM&174,776&ARM-PPC&174,916\\
x86-PPC&182,596&ARM-x64&157,976\\
x86-x64&167,436&PPC-x64&164,916\\
\hline
\end{tabular}
}
\caption{Number of function pairs for model training and testing. Arch-Comb means architecture combination.}
\label{tab:funcpair}\vspace{-15pt}
\end{table}

\subsection{Baseline Approaches}
There have been several previous works for BCSD study: \textit{discovRE}~\cite{eschweiler2016discovre}, \textit{Genius} \cite{feng2016scalable},  \gemini{} \cite{xu2017neural}, and \diaphora{} \cite{diaphora}. 
Xu \etal{} have demonstrated that \gemini{} is both more accurate and efficient than the other approaches~\cite{xu2017neural}.
In our evaluation, \gemini{} is chosen as one of the baseline approaches.
Considering that \diaphora{} \cite{diaphora} also uses the AST as the features for similarity calculation and is not compared with other works, we also choose it as a baseline for comparison.
\paragraph*{\textbf{Gemini}}
\gemini{} encodes ACFGs (attributed CFGs) into vectors with a graph embedding neural network.
The ACFG is a graph structure where each node is a vector corresponding to a basic block.
We have obtained \gemini{}'s source code and its training dataset.
Notice that in~\cite{xu2017neural} authors mentioned it can be retrained for a specific task, such as the bug search.
To obtain the best accuracy of \gemini{}, we first use the given training dataset to train the model to achieve the best performance.
Then we re-train the model with the part of our training dataset.
For the \gemini{} evaluation, we construct ACFG pairs from the same function pairs with AST pairs in our OpenSSL dataset.

\paragraph*{\textbf{Diaphora}}
We download the \diaphora{} source code from github \cite{diaphora}.
\diaphora{} maps nodes in an AST to primes and calculates the product of all prime numbers.
Then it utilizes a function to calculate the similarity between the prime products.
We extract \diaphora{}'s core algorithm for AST similarity calculation for comparison.

\subsection{Evaluation Metric}\label{evaluation}
The Receiver Operating Characteristic \ysgbf{(ROC)} curve and Area Under Curve \ysgbf{(AUC)} are used for measuring the model performance.
The ROC curve \cite{zweig1993receiver} illustrates the diagnostic ability of a model as its discrimination threshold is varied.
For our evaluation, the AUC reflects the probability that the model will correctly identify whether an AST pair is from a homologous pair or not.
In our evaluation, the similarity of a function pair is calculated as a score of $r$. 
Assuming the threshold is $\beta$, if the similarity score $r$ of a function pair is greater than or equal to $\beta$, the function pair is regarded as a positive result, otherwise a negative result.
For a homologous pair, if its similarity score $r$ is greater than or equal to $\beta$, it is a true positive \ysgbf{(TP)}. If a similarity score of $r$ is less than $\beta$, the calculation result is a false negative \ysgbf{(FN)}. 
For a non-homologous pair, if a similarity score $r$ is greater than or equal to $\beta$, it is a false positive \ysgbf{(FP)}.
When the similarity score $r$ is less than $\beta$, it is a true negative \ysgbf{(TN)}. 
After classifying all the calculation results, two important metrics \ysgbf{$TPR$} (true positive rate) and \ysgbf{$FPR$} (false positive rate) under certain threshold $\beta$ are calculated as:
    \begin{equation}
    \begin{aligned}
    TPR = \frac{TP}{TP+FN}
    \end{aligned}
    \end{equation}
        \begin{equation}
    \begin{aligned}
    FPR = \frac{FP}{FP+TN}
    \end{aligned}
    \end{equation}
The \ysgbf{ROC} curve can be generated by plotting points whose coordinates consist of \ysgbf{$FPR$}s and \ysgbf{$TPR$}s with many different thresholds.
After ROC curves being plotted, the area under the \ysgbf{ROC} curve, called \ysgbf{AUC} for short, can be calculated.
The larger the \ysgbf{$AUC$}, the better the model's discrimination ability.

\subsection{Evaluation Results}

We conduct comparative experiments to assess the performances of three approaches: \ProjectName{}, \toolname{Gemini} and \toolname{Diaphora}.
To measure the performance gain of the calibration scheme described in~\cref{sec:simcor}, we also test the performance of \ProjectName{} without calibration (\ie, \ProjectName{}\toolname{-WOC}), where the calibration algorithm is not included and the AST similarity is directly used as the final function similarity.
We conduct two comparative experiments: the mixed cross-architecture experiment and the pair-wise cross-architecture experiment.
In the mixed cross-architecture experiment, the function pairs are randomly constructed from any architecture combinations.
In particular, the functions in such function pairs could come from any platform of x64, x86, PPC, or ARM.
The pair-wise cross-architecture experiments aim to compare the performance in a specific architecture combination.
Therefore, six different function combinations are constructed: \textit{ARM-PPC, ARM-x64, PPC-x64, x86-ARM, x86-PPC}, and \textit{x86-x64}, respectively.
In each individual combination, function pairs are selected from two specific different architectures.
For example, in the \textit{ARM-PPC} cross-architecture experiment, the functions in the pairs only come from the ARM or PPC architecture.

\begin{figure}[tb]
\centering
  \centering
   \includegraphics[scale=0.38]{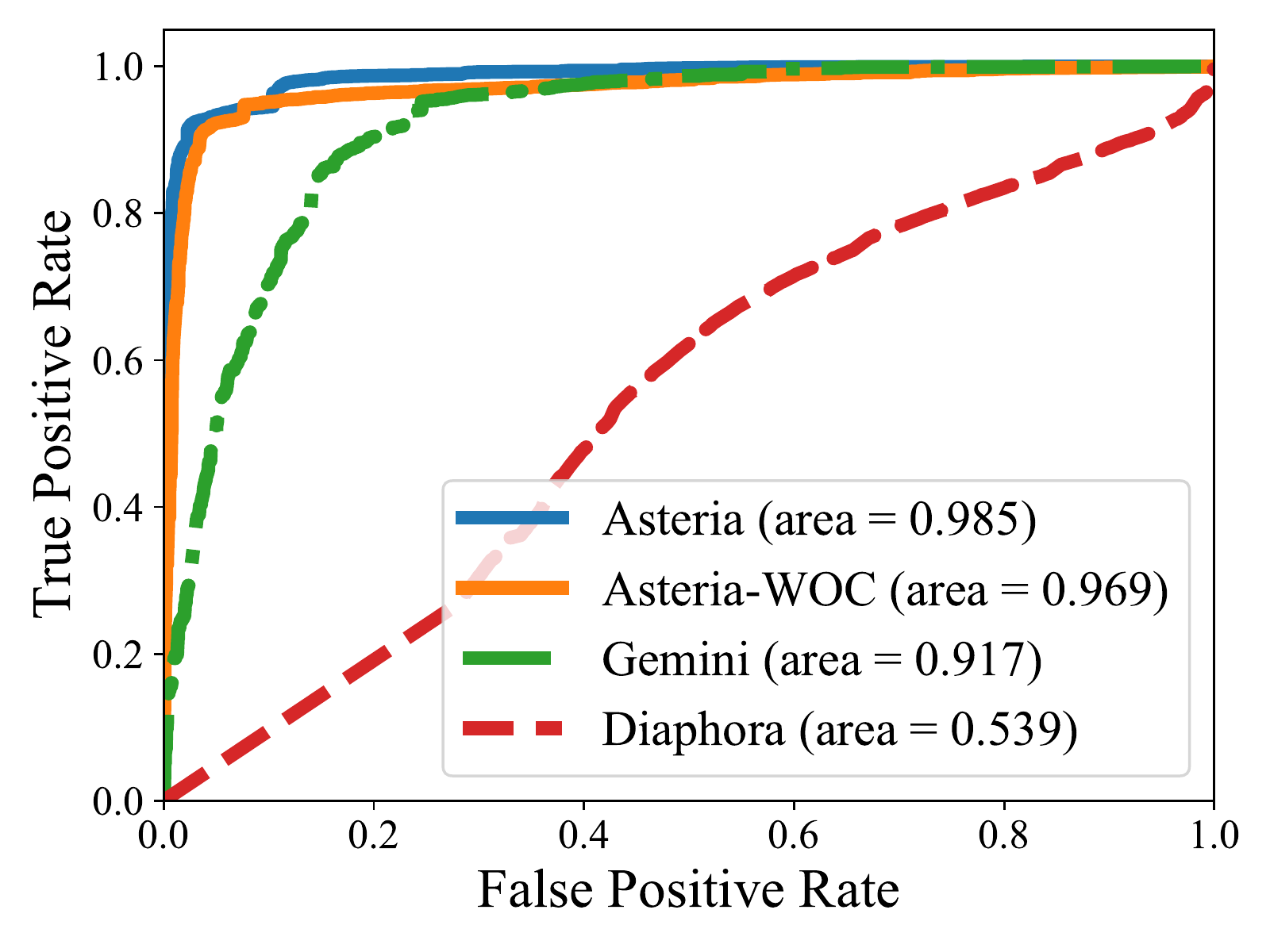}\vspace{-10pt}
  \caption{The ROC curves for \ProjectName{}, \ProjectName{}\toolname{-WOC}, \gemini{}, and \diaphora{} in mixed cross-architecture evaluation}
  \label{fig:rocs}
\end{figure}%
\begin{figure}[tb]
\vspace{-10pt}
  \centering
  \includegraphics[scale=0.40]{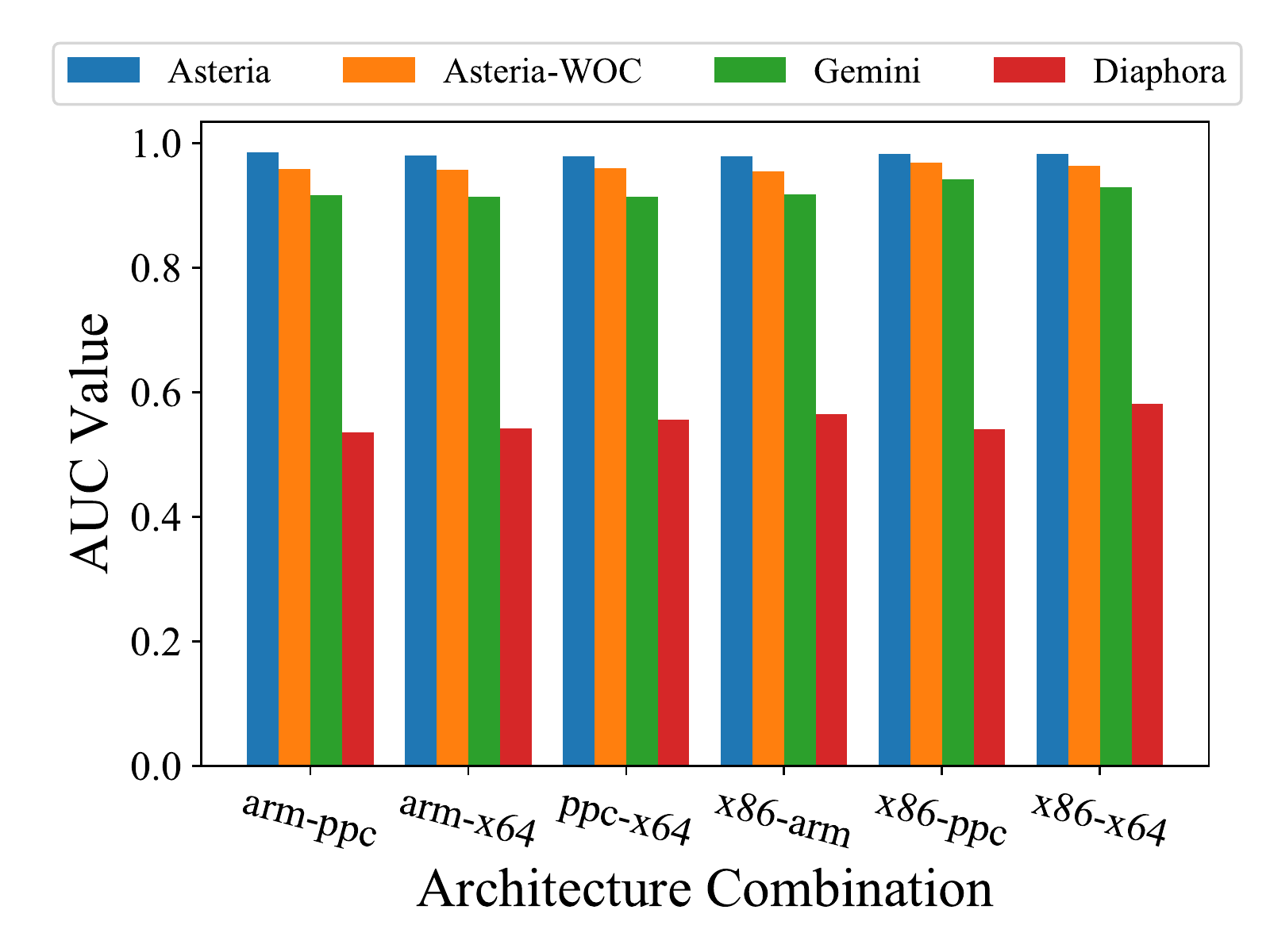}\vspace{-10pt}
 \caption{The AUCs for \ProjectName{}, \ProjectName{}\toolname{-WOC}, \gemini{}, and \diaphora{} in pair-wise cross-architecture evaluation.}\vspace{-18pt}
 \label{fig:aucs}
\end{figure}
\subsubsection{ROC Comparison}\label{sec:roc}

Figure \ref{fig:rocs} plots the ROC curves of the different approaches in cross-architecture experiments.
The result shows that \ProjectName{} outperforms \gemini{} by around 7.5\% and \diaphora{} by 82.7\%, and the similarity detection accuracy of \ProjectName{} has been enhanced with the similarity calibration (\cref{sec:simcor}).
Note that our approach achieves a high true positive rate with a low false positive rate, which brings out high confidence of calculation results in practical applications.
For example, from Figure \ref{fig:rocs}, with a low false-positive of 5\%, our approach achieves a true positive rate of 93.2\%, while Gemini has a true positive rate of 55.2\%.
As shown in Figure \ref{fig:aucs}, we calculate the AUC values of different approaches in six different pair-wise experiments.
The results show that \ProjectName{} outperforms \gemini{} and \diaphora{} in each pair-wise architecture combination by similar margins as in the mixed cross-architecture experiment.
The major difference between \ProjectName{} and \gemini{} is that we take AST as our function feature, which shows that AST is better than CFG as a semantically related feature.
Compared to \diaphora{}, the results show that the application of NLP technology greatly improves the performance of AST based BCSD.

\subsubsection{Impact of Model Settings}\label{ModelSettings}
We illustrate the impact of four hyperparameters on the model performance: epochs, siamese structure, embedding size, and leafnode calculation. 
The experimental settings are the same as those described in Section \ref{sec:modelsetting} during evaluation except for the parameters being measured.

%
\begin{figure}[tb]
    \centering
    \includegraphics[scale=0.38]{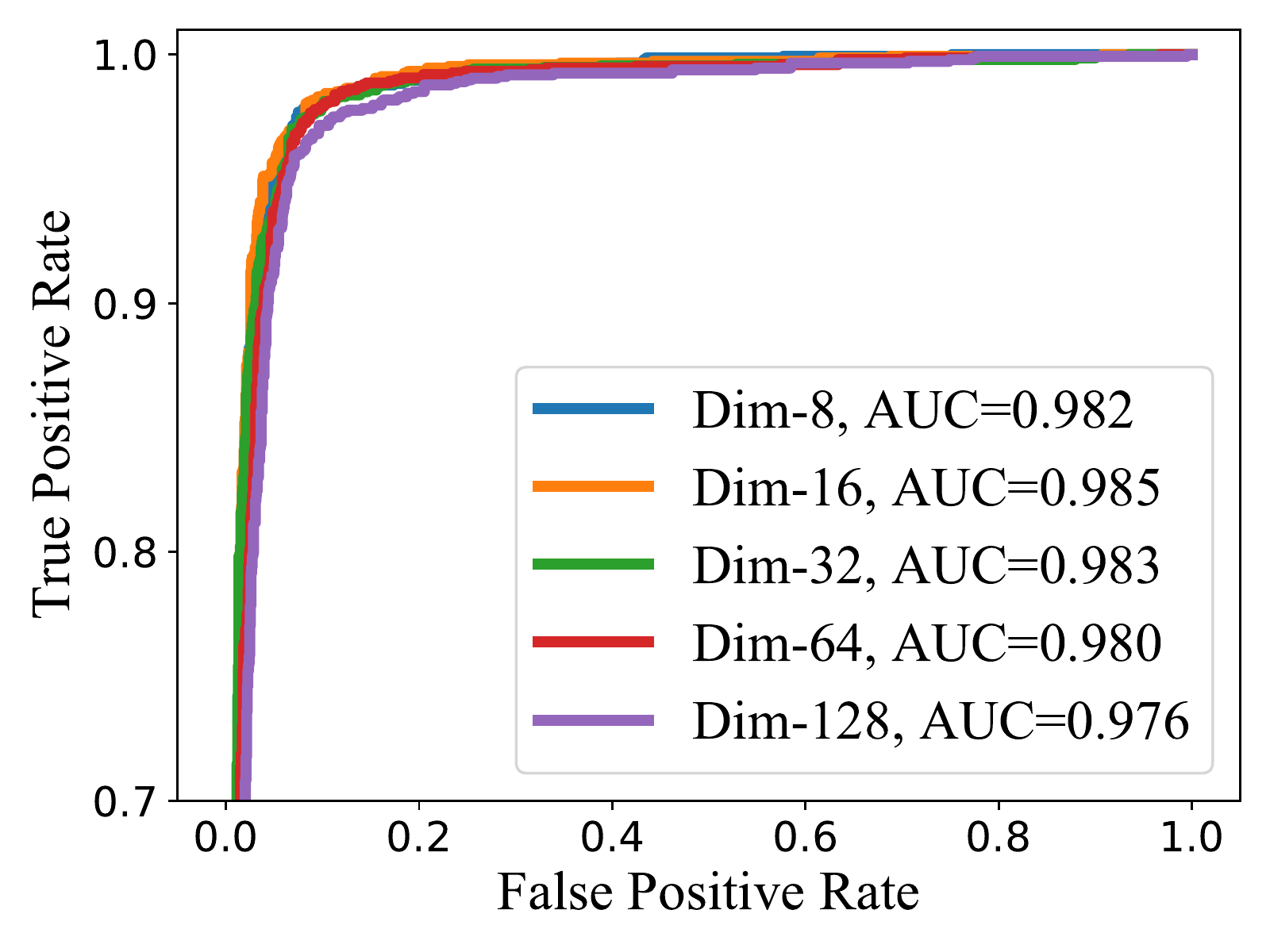}
    \vspace{-10pt}
    \caption{ROC curves in different embedding sizes from $8$ to $128$.\label{fig:emb_size}}
    \vspace{-16pt}
\end{figure}

\begin{figure}[tb]
    \centering
    \includegraphics[scale=0.38]{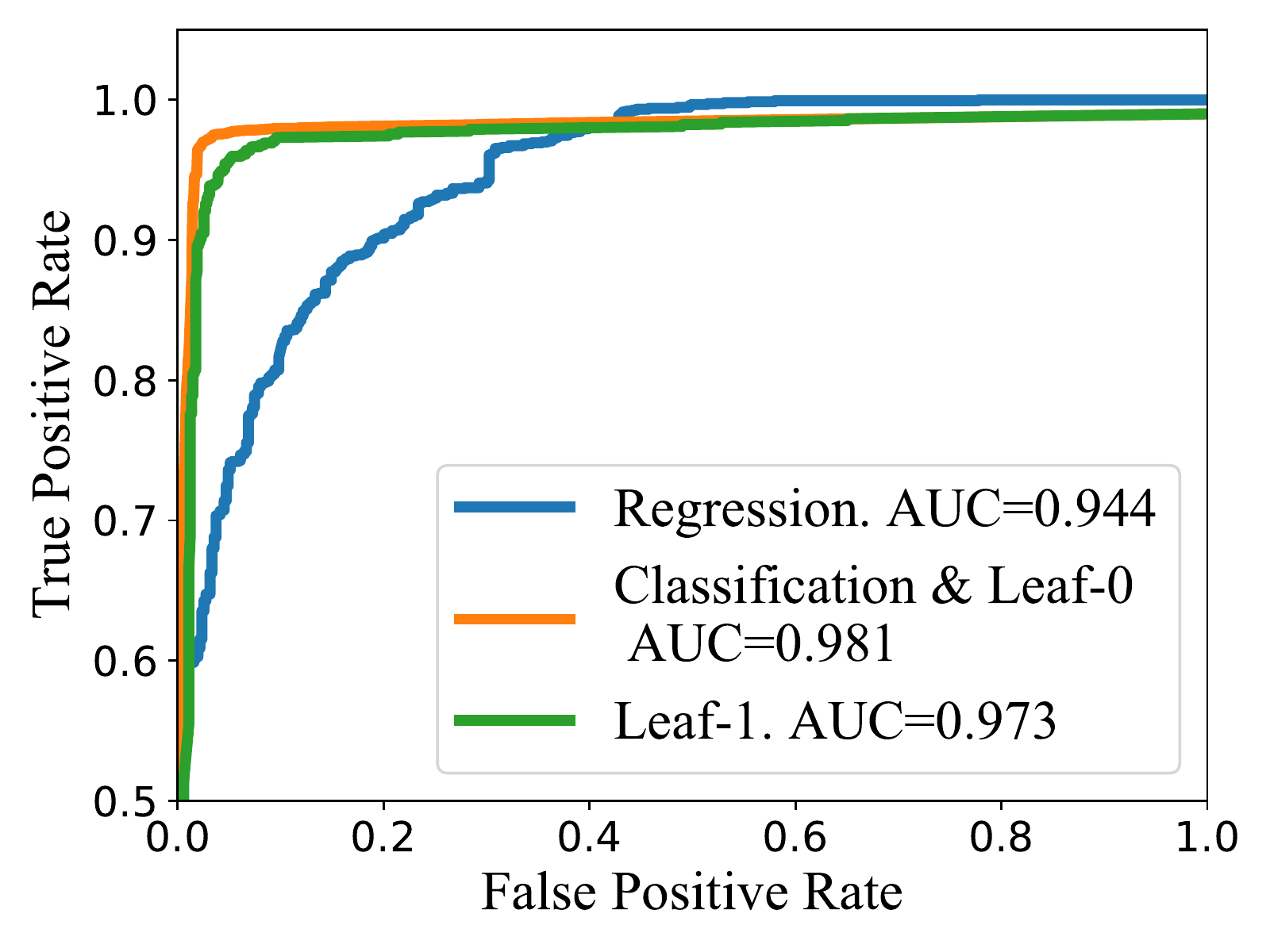}  
    \caption{Impact of Siamese structures and leaf node calculation.\label{fig:pro_trans}}
    \vspace{-18pt}
\end{figure}

\begin{figure*}[t]
    \centering
    \begin{minipage}[c]{0.33\textwidth}
    \centering
    \includegraphics[scale=0.36]{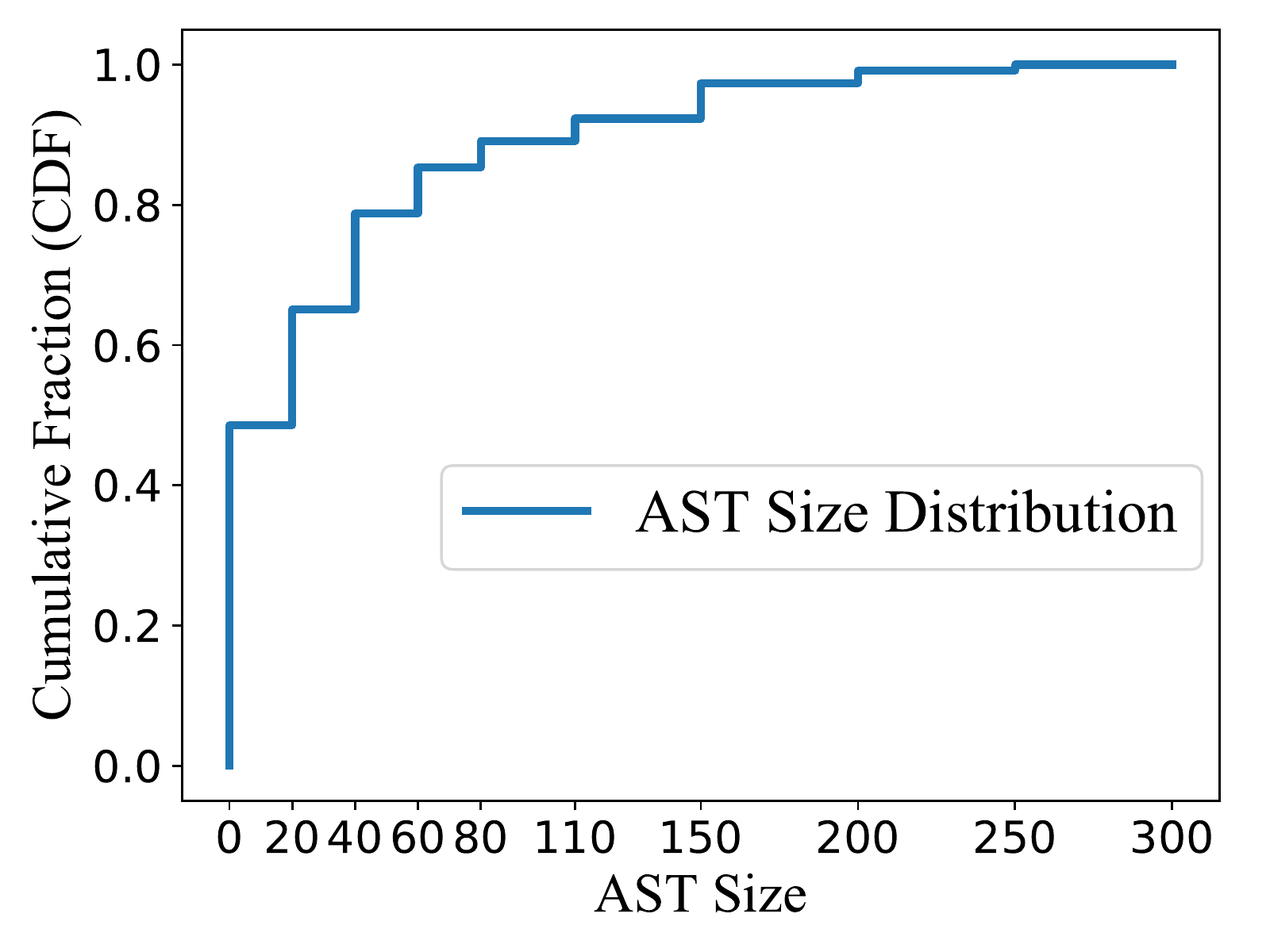}
    \end{minipage}
     \begin{minipage}[c]{0.33\textwidth}
    \centering
    \includegraphics[scale=0.36]{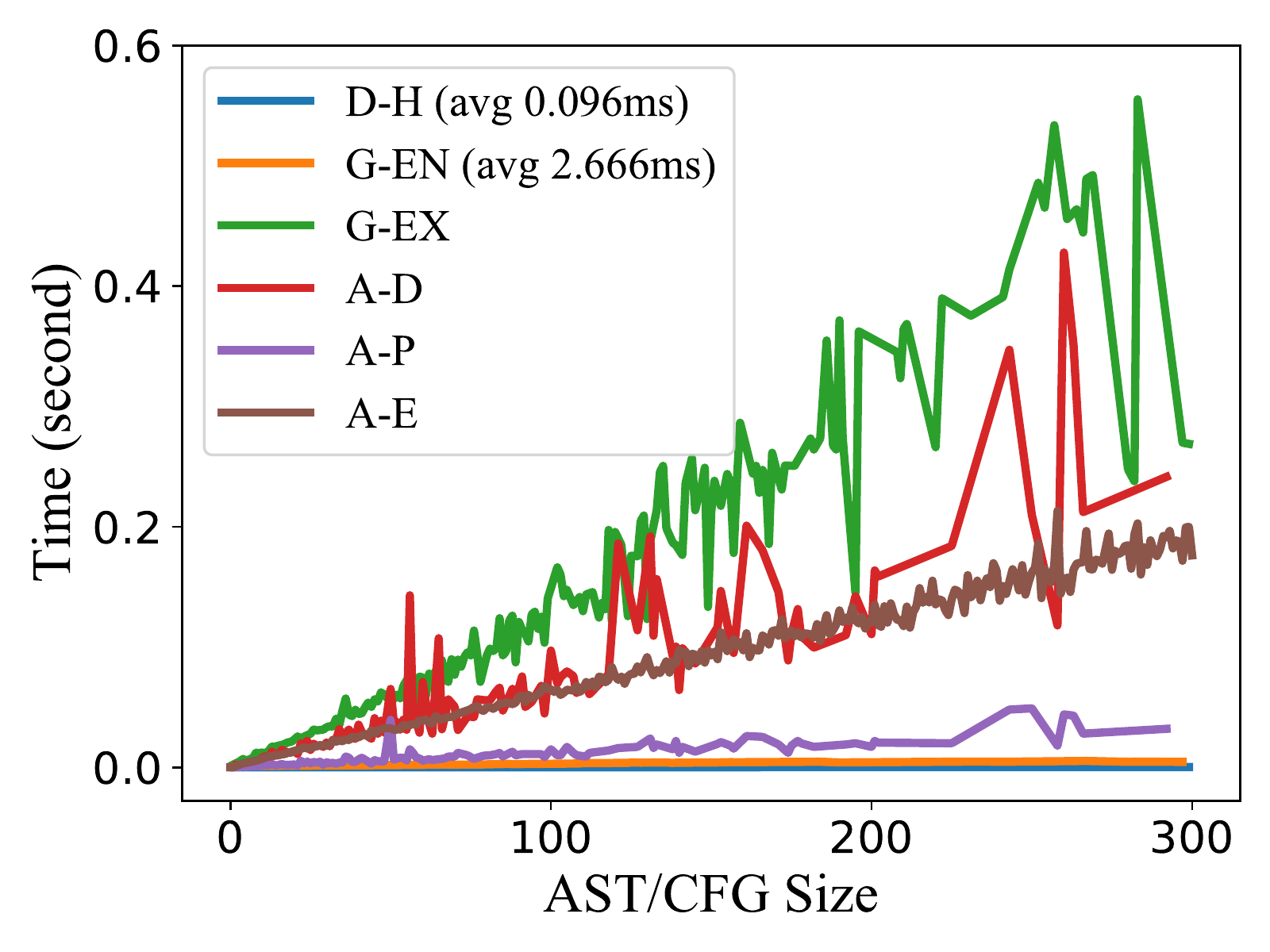}
    \end{minipage}%
    \begin{minipage}[c]{0.33\textwidth}
    \centering
    \includegraphics[scale=0.36]{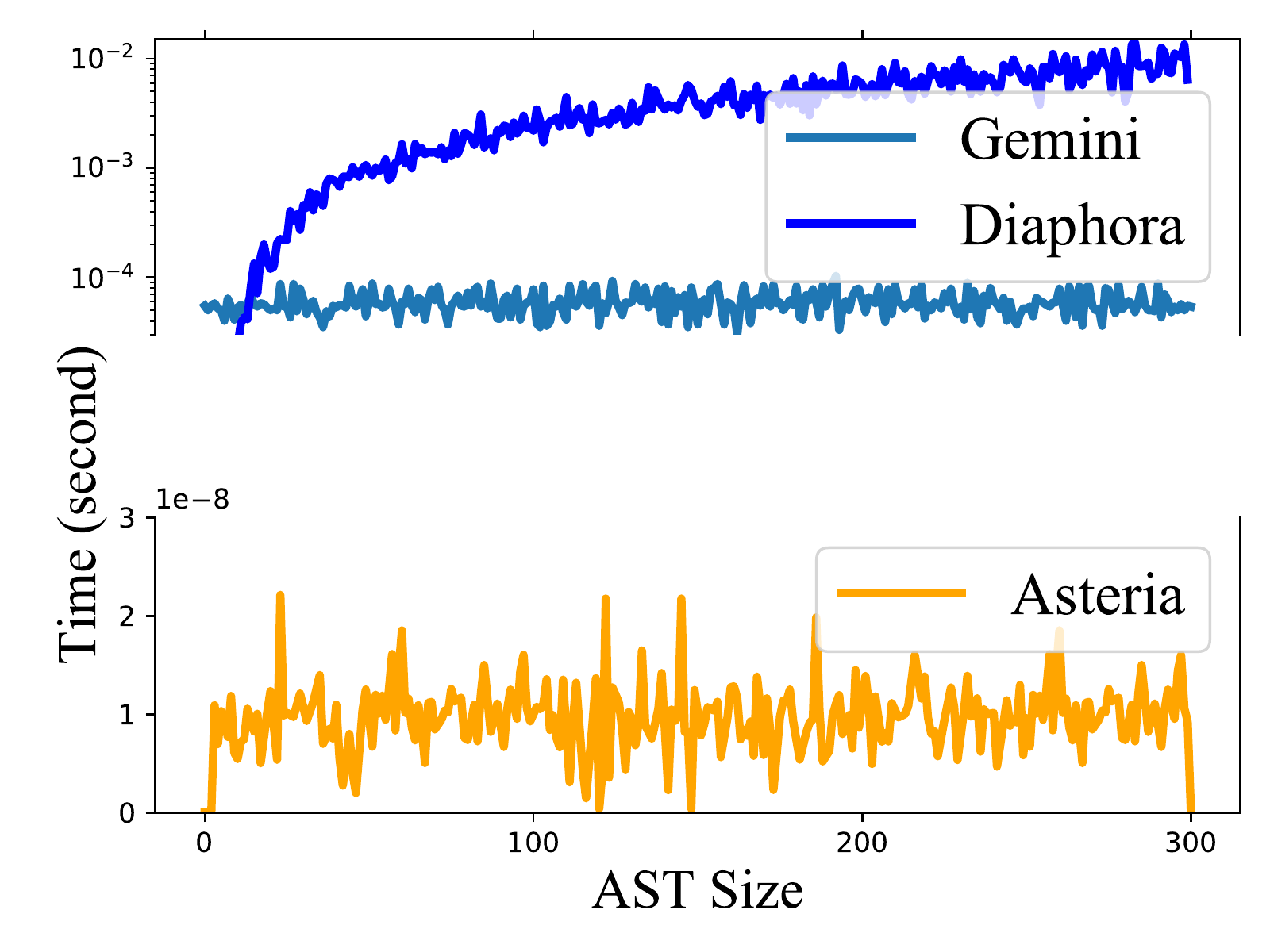}   
    \end{minipage}
     \begin{minipage}[t]{0.32\textwidth}
     \vspace{-3pt}
    \subcaption{Cumulative distribution of AST size for 25,569 ASTs.\label{fig:sourcecode}}
    \end{minipage}
    \hfill
    \begin{minipage}[t]{0.32\textwidth}
    \vspace{-3pt}
    \subcaption{Times overhead in the offline phases of \ProjectName{}, Gemini, and Diaphora.
    \label{fig:dectime}}
    \end{minipage}
    \hfill
     \begin{minipage}[t]{0.32\textwidth}
     \vspace{-3pt}
    \subcaption{Time overhead in the online phases of different approaches. \label{fig:timecost}}
    \end{minipage}
    \vspace{-8pt}
    \caption{Computational overhead.}
\end{figure*}

\paragraph{\textbf{Embedding Size}}
The embedding size is the dimension of the vector outputted by the embedding layer during the AST encoding (\cref{sec:encoding}).
To show the difference of model performance with different embedding sizes, we increase the embedding size from $8$ to $128$ for model training and testing as shown in Figure~\ref{fig:emb_size}.
We take the highest AUC value from the model testing in each embedding size setting.
The results show that the model achieves the highest AUC of $0.985$ with the embedding size of 16, and reaches the lowest AUC of $0.976$ with the embedding size of 128.
The embedding size of 128 used to represent 43 kinds of nodes in Table~\ref{tab:mapping} increases the model complexity and may cause the overfitting problem.
Considering the tradeoff between model performance (AUC) and computational complexity, we choose the embedding size of 16 in \ProjectName{}.

\paragraph{\textbf{Siamese Structure}}\label{sec:siamesestructure}
The similarity calculation between AST encoding vectors in \siamese{} is introduced in \cref{sec:modeldesc}.
There is another common way to calculate the similarity between two encoding vectors called the regression method, which utilizes the cosine distance~\cite{url:cosdistance}.
Different similarity calculation methods correspond to different \siamese{} internal structures.
In contrast, we adopt a calculation method similar to the binary classification as shown in Equation~(\ref{equation:classification}) in \ProjectName{}.
For comparison with the regression method, we use the same model settings except for the similarity calculation step.
We record the $TPR$s and $FPR$s and plot the ROC curve corresponding to the different internal structures.
As Figure \ref{fig:pro_trans} shows, the curve with the label ``Regression'' corresponds to the calculation with the cosine distance, and the curve with the label "Classification" corresponds to the calculation in \ProjectName{}.
The result shows that the model with our calculation method yields a higher AUC value of $0.981$, which means a better model performance.

\paragraph{\textbf{Leafnode Calculation}}\label{sec:leafnodeinitialization}
According to Equation (\ref{equation:ck}) in \cref{sec:modeldesc}, the calculation for leaf nodes also takes the hidden state input and cell state input. 
Since leaf nodes have no children, we assign the input vectors of the leaf nodes manually.
There are two ways for vector initialization in general: all zeros vector or all ones vector.
According to these two different initialization methods, we train and test the model, and then plot the ROC curves.
As shown in Figure~\ref{fig:pro_trans}, the curve with label ``Leaf-0'' corresponds to the model with all zeros initialization, and the curve with label ``Leaf-1'' corresponds to the model with all ones initialization.
The results show the model performs better with all zeros initialization. 

\subsection{Computational Overhead} 
\subsubsection{AST Statistics}
We define the AST size as the number of nodes in an AST.
Since there are global offset table functions~\cite{cabrero2001system} which are automatically generated by the compiler, we filter out such functions.
25,569 ASTs are retained after filtering from 25,667 ASTs.
We sort the sizes of 25,569 AST samples in ascending order from the OpenSSL dataset.
Figure~\ref{fig:sourcecode} plots the cumulative distribution of AST size of the sorted AST samples.
It shows that most AST sizes are smaller than 200. 
Specifically, the ASTs with sizes less than 20, 40, 80, and 200 account for 48.6\%, 65.1\%, 85.4\%, and 97.4\%, respectively.
{We then measure the computational overhead of AST extracting and encoding with \ProjectName{} and similarity calculation with different approaches.}

\subsubsection{Time Consumption}
The process of function similarity calculation consists of two phases: offline phase (AST extraction, preprocessing, and encoding) and online phase (similarity calculation between encoding vectors).
We separately measure the extraction time, preprocessing time, encoding time, and similarity calculation time on the {\bf OpenSSL dataset}.
The decompilation is the most time-consuming part in the AST extraction.
The average time for each function decompilation is around 41 milliseconds, and the average time for preprocessing is around 6 milliseconds. 
We only consider ASTs whose sizes are less than 300 since they account for the majority of ASTs, as illustrated in Figure~\ref{fig:sourcecode}. 
Figure~\ref{fig:dectime} shows the encoding time of ASTs, which is about the same as the decompilation time.
In the figure, D-H denotes the AST hashing time of Diaphora. 
G-EN denotes the ACFG encoding time of Gemini and G-EX denotes the ACFG extraction time.
A-D, A-P, A-E respectively represent the decompilation time, preprocessing time, encoding time of \ProjectName{}.
\ProjectName{} has a more time-consuming offline stage than Gemini and Diaphora.
The similarity calculation time in the online phase is shown in Figure~\ref{fig:timecost}.
\ProjectName{} takes $8 \times 10^{-9}$ second on average to calculate the similarity for a pair of encoded ASTs. However, the average calculation time of \diaphora{} and \gemini{}  is $4 \times 10^{-3}$ second and $6 \times 10^{-5}$ second for a pair of ASTs, respectively.
Remarkably, it shows that \ProjectName{} is much faster than \diaphora{} and \gemini{} in terms of the similarity calculation. The root cause of better performance is that we use the subtraction and production of vectors for the similarity calculation, while Gemini uses the cosine function which is much more time-consuming.

\begin{table*}[h]
    \centering
	\resizebox{18.0cm}{!}{
\begin{tabular}{ccccp{1.9cm}cc}
\hline
\rowcolor[HTML]{C0C0C0}  \# & CVE & software version & vulnerable function & \textbf{Confirmed \#} & Vendor & Affected model\\\hline
\multirow{2}*{1} & \multirow{2}*{2016-2105~\cite{url:CVE-2016-2105}} & \multirow{2}*{\shortstack{OpenSSL before 1.0.1t \\and 1.0.2 before 1.0.2t}} & \multirow{2}*{EVP\_EncodeUpdate} & \multirow{2}*{\textbf{24}} & \multirow{2}*{NetGear} & \multirow{2}*{\shortstack{D7000, R7000, FVS318Gv2, R8000, R7500,\\R7800, R6250, R7900, R7900, R7800}}\\
\\
\hline
2 & 2014-4877~\cite{url:CVE-2014-4877} & \shortstack{Wget before 1.16} & ftp\_retrieve\_glob & \textbf{14} & NetGear &R7000, D7800, R7800, R6250, R8000, R7900, R6700 \\
\hline
\multirow{2}*{3} & \multirow{2}*{2014-0195~\cite{url:CVE-2014-0195}} & \multirow{2}*{\shortstack{OpenSSL before 0.9.8za, 1.0.0 before\\ 1.0.0m, and 1.0.1 before 1.0.1h}} & \multirow{2}*{dtls1\_reassemble\_fragment} & \multirow{2}*{\textbf{13}} & NetGear
&R7000, R8000, R6250, R7900, R7500, D7800 \\
&&&&&Dlink&DSN-6200\\
\hline
4 & 2016-6303~\cite{url:CVE-2016-6303} & OpenSSL before 1.1.0 &  MDC2\_Update & \textbf{13} & NetGear
&R7000, D7000, FVS318Gv2, R8000 \\
\hline
5 & 2016-8618~\cite{url:CVE-2016-8618} & Libcurl before 7.51.0 &  curl\_maprintf & \textbf{7} & NetGear
&D7000, R8000, R6700, R7500, D7800, R7800\\
\hline
6 & 2013-1944~\cite{url:CVE-2013-1944} & Libcurl before 7.30.0 &  tailmatch & \textbf{3} & NetGear
&R7500, D7800, R7800\\
\hline
7 & 2011-0762~\cite{url:CVE-2011-0762} & vsftpd before 2.3.3&  vsf\_filename\_passes\_filter & \textbf{1} & NetGear
&FVS318Gv2\\
\hline 
\end{tabular}
}
\caption{Results of vulnerability search. ``\textbf{Confirmed \#}'' indicates the number of confirmed vulnerable functions.}
\label{tab:application}    \vspace{-10pt}
\end{table*}
\section{Vulnerability Search}
We conduct a vulnerability search using the {\bf Firmware dataset} and successfully find a total of \textbf{75} vulnerabilities. We first encode all the functions in the Firmware dataset with the Tree-LSTM network and obtain the encoding vectors and the number of callee functions.
We build a vulnerability library that contains vulnerable functions exposed in commonly used open-source software. 
In the library, seven vulnerable functions of CVE are integrated and the details of vulnerable functions are listed in Table \ref{tab:application}. 
Specifically, three of seven come from OpenSSL software, and two of them come from \textit{libcurl}, and one comes from \textit{wget}, and the other comes from \textit{vsftpd}.
For each vulnerable function, we also encode it with the Tree-LSTM network, and record the encoding vector and the number of callee functions. 
Then we perform the similarity calculation between the encoding vectors of vulnerable functions and the encoding vectors of all functions in the Firmware database with the \siamese{}.
After the calculation, each function $f$ in the Firmware dataset is tagged with a similarity score of $r_{f,v}$ with the vulnerable function $v$. 
According to the Youden Index~\cite{youden1950index}, which measures model performance at a certain threshold.
Since our model performs best at the threshold value of $0.84$, we set the threshold to be $0.84$ to filter out $924$ candidate functions.
A score of $r_{f, v}$ greater than the threshold indicates that function $f$ may be a homologous vulnerable function of $v$.
If $f$ is considered a vulnerable function, we further conduct a manual analysis to confirm whether it is indeed a homologous function of $v$ or not.
In our analysis, we define two criteria: \textbf{A}) function $f$ comes from the same version of the same software as the vulnerable function $v$, and \textbf{B}) the score of $r_{f,v}$ is 1.
The candidate functions that satisfy the criteria A and B are considered as homologous functions of the corresponding vulnerable functions with high confidence, so that we won't conduct further manual analysis. 
For those candidate functions that satisfy the criteria A but not B,  we further analyze the semantic functionality of their assembly code to determine whether they are homologous functions of the corresponding vulnerable functions.
From the search results, 68 vulnerable functions satisfy criteria A, 13 vulnerability functions satisfy criteria B, and 6 functions satisfy A and B.
The number of homologous vulnerable functions corresponding to each vulnerable function of \ysgbf{CVE} is listed in Table \ref{tab:application}.
Since a device model corresponds to multiple versions of firmware and different versions likely hold the same vulnerability, the number of the device models is less than the number of confirmed vulnerable functions.

We also conduct a comparison between \ProjectName{} and \gemini{} in terms of the end-to-end time consumption and accuracy for the vulnerable function search.
We compare the end-to-end time consumption and accuracy for the top 10 vulnerable functions in the search results.
\ProjectName{} takes 0.414 seconds on average for a pair of functions and achieves 78.7\% accuracy.
The \gemini{} takes 0.159 seconds for a pair and achieves 20\% accuracy.
By checking the results of \gemini{}, we find that most of the vulnerable functions being ranked outside the top 10000 functions, which means a high false negative rate of results. 
In comparison, \ProjectName{} produces less false negatives, which means \ProjectName{} is more effective than \gemini{}.

\section{Related Work}


\paragraph{\textbf{Feature-based Methods}}
When considering the similarity of binary functions, the most intuitive way  is to utilize the assembly code content to calculate the edit distance for similarity detection between functions.
Khoo \etal{} concatenated consecutive mnemonics from assembly language into the N-grams for similarity calculation~\cite{khoo2013rendezvous}.
David \etal{} proposed Trecelet, which concatenates the instructions from adjacent basic blocks in CFGs for similarity calculation~\cite{David:Tracelet}.
Saebjornsen \etal{} proposed to normalize/abstract the operands in instructions, \eg{} replacing registers such as eax or ebx with string ``reg'', and conduct edit distance calculation based on normalized instructions~\cite{Saebjornsen:2009}.
However, binary code similarity detection methods based on disassembly text can not be applied to cross-architecture detection since the instructions are typically different in different architectures.
The works in \cite{pewny2015cross}, \cite{chandramohan2016bingo}, \cite{eschweiler2016discovre}, \cite{xue2018accurate} utilize cross-architecture statistical features for binary code similarity detection.
Eschweiler \etal{}~\cite{eschweiler2016discovre} defined statistical features of functions such as the number of instructions, size of local variables.
They utilized these features to calculate and filter out candidate functions.
Then they performed a more accurate but time-consuming calculation with the graph isomorphism algorithm based on CFGs.
Although this method takes a pre-filtering mechanism, the graph isomorphism algorithm makes similarity calculation extremely slow.
To improve the computation efficiency, Feng \etal{} proposed {Genius} which utilizes machine learning techniques for function encoding~\cite{feng2016scalable}.
{Genius} uses the statistical features of the CFG proposed in~\cite{eschweiler2016discovre} to compose the attributed CFG (ACFG).
Then it uses a clustering algorithm to calculate the center points of ACFGs and forms a codebook with the center points.
Finally, a new ACFG is encoded into a vector by computing the distance with ACFGs in the codebook and the similarity between ACFGs is calculated based on the encoded vectors.
But the codebook calculation and ACFG encoding in {Genius} are still inefficient.
Xu \etal{} proposed \gemini{} based on Genius to encode ACFG with a graph embedding network~\cite{xu2017neural} for improving the  accuracy and efficiency.
However, the large variance of binary code across different architectures makes it difficult to find architecture-independent features~\cite{d2015correctness}.

\paragraph{\textbf{Semantic-based Methods}}
For more accurate detection, semantic-based features are proposed and applied for code similarity detection.
The semantic-based features model the code functionality, and are not influenced by different architectures.
Khoo \etal{} applied symbolic execution technique for detecting function similarity~\cite{Luo:SoftwarePlagiarism}.
Specifically, they obtained input and output pairs by executing basic blocks of a function.
But the input and output pairs can not model the functionality of the whole function accurately.
Ming \etal{} leveraged the deep taint and automatic input generation to find semantic differences in inter-procedural control flows for function similarity detection~\cite{ming2012ibinhunt}.
Feng~\etal{} proposed to extract conditional formulas as higher-level semantic features from the raw binary code to conduct the binary code similarity detection~\cite{feng2017extracting}.
In their work, the binary code is lifted into a platform-independent intermediate representation (IR), and the data-flow analysis is conducted to construct formulas from IR.
Egele \etal{} proposed the blanket execution, a novel dynamic equivalence testing primitive that achieves complete coverage by overwriting the intended program logic to perform the function similarity detection~\cite{BlanketExecution}.
These semantic-based features capture semantic functionalities of a function to reduce the false positives.
However, the methods above depend heavily on emulation or symbolic execution, which are not suitable for program analysis in large-scale IoT firmware since the emulation requires peripheral devices~\cite{zaddach2014avatar, gustafson2019toward, chen2016towards} and symbolic execution suffers from the problems of path explosion.

\paragraph{\textbf{AST in Source Code Analysis}}
\clnote{Since the AST can be easily generated from source code, there has been research work proposed to detect source code clone based on AST.
Ira D. Baxter \etal{} proposed to hash ASTs of functions to buckets and compare the ASTs in the same bucket ~\cite{baxter1998clone} to find clones.
Because the method proposed in~\cite{baxter1998clone} is similar to \diaphora{} which hash ASTs, we only perform a comparative evaluation with \diaphora{}.
In addition to the code clone detection, AST is also used in vulnerability extrapolation from source code~\cite{yamaguchi2012generalized,yamaguchi2011vulnerability}. In order to find vulnerable codes that share a similar pattern, Fabian \etal{}~\cite{yamaguchi2012generalized} encoded AST into a vector and utilized the latent semantic analysis~\cite{deerwester1990indexing} to decompose the vector to multiple structural pattern vectors and compute the similarity between these pattern vectors.}
Yusuke Shido~\etal{} proposed an automatic source code summary method with extended Tree-LSTM~\cite{shido2019automatic}.

\section{Discussion}
In this section, we discuss the limitation of this work and future research to address these limitations. 

Our work focuses on a more accurate and faster method for cross-platform binary code similarity detection.
In \ProjectName{}, we use AST as a feature to capture semantic information of a function.
Considering the complexity of decompilation in multiple platforms, we use the commercial tool \texttt{IDA Pro}.
\clnote{Our work relies on the correctness of the decompilation result of \texttt{IDA Pro}.}

For the digitization of nodes before encoding an AST into the Tree-LSTM network, we remove the constant values and strings in the AST since they correspond to unlimited instances. 
Without the constant values and strings, the semantic information of a function may be relatively incomplete and may cause false positives. For the more accurate semantic representation of a function, we may introduce another embedding system to embed constants and strings into embedding vectors, and combine the embedding vectors with the AST encoding in \ProjectName{} to get a new AST representation. 
There is no doubt that adding a new embedded system will increase the computational overhead. 
But it achieves a good tradeoff between the computational cost and accuracy.

For the organization strategy of function pairs introduced in Section~\ref{dataset}, it may introduce few noises in case two functions with different names have the exactly same code logic within the same project (despite rarely in practice based on our observation). Considering the small probability of data noise and the tolerance of Tree-LSTM to data noise, we believe that our method is minimally affected.
In terms of the computational overhead, \ProjectName{} is based on the Tree-LSTM network, where the model training is time-consuming if the size of an AST is large.
In addition, the offline encoding is time-consuming since LSTM cannot take full advantage of GPU for parallel computing.
As our future research, we plan to employ more efficient NLP techniques for AST encoding.

We conduct a vulnerability search in our firmware dataset and analyze the search results.
Through our analysis, we find that the results with score of 1 are all confirmed as vulnerable cases. The functions with high scores hold similar code patterns with the vulnerable function.
We hope that this work can inspire researchers about the connection between natural language and assemble language or intermediate language for binary security analysis.
Notice that vulnerability verification is a very important follow-up work after the vulnerability search. We only manually confirm whether the functions are vulnerable or not, without conducting the exploitability analysis.

\section{Conclusion}

In this paper, we proposed \ProjectName{}, a deep learning-based AST encoding scheme to measure the semantic equivalence of functions across platforms. 
We used AST as the function feature and adopted the Tree-LSTM network to encode the AST into semantic representation vectors.
We then employed the \siamese{} for integrating two identical Tree-LSTM networks to calculate the similarity between two AST encodings.
We implemented a prototype of \ProjectName{}. For the model training, we built a large-scale cross-platform dataset containing 49,725 binary files by cross-compiling 260 open-source software. 
We compared our model against the state-of-the-art approach \gemini{} and an AST-based method \diaphora{}. Comprehensive evaluation results show that our \ProjectName{} outperforms \diaphora{} and \gemini{} in both accuracy and efficiency.
We also conducted a vulnerability search in an IoT firmware dataset and found 75 vulnerable functions.

\section{Acknowledgement}
This work is partly supported by Guangdong Province Key Area R\&D Program of China (Grant No.2019B010137004), Key Program of National Natural Science Foundation of China (Grant No.U1766215), National Natural Science Foundation of China (Grant No.U1636120) and Industrial Internet Innovation and Development Project (Grant No.TC190H3WU).

\bibliographystyle{plain}
\bibliography{ref.bib}

\end{document}